\documentclass[twocolumn,epjc3]{svjour3}

\RequirePackage{graphicx}

\RequirePackage{amsmath,amsfonts,amssymb}
\RequirePackage[numbers,sort&compress]{natbib}
\RequirePackage[colorlinks,citecolor=blue,urlcolor=blue,linkcolor=blue]{hyperref}
\RequirePackage{dsfont}

\usepackage[utf8]{inputenc}

\journalname{Eur. Phys. J. C}

\begin{document}
\title{Starobinsky cosmological model in Palatini formalism}

\author{Aleksander Stachowski\thanksref{oauj,e-as}
\and
Marek Szyd{\l}owski\thanksref{oauj,csrc,e-ms}
\and
Andrzej Borowiec\thanksref{itpwu,e-ab}}

\thankstext{e-as}{aleksander.stachowski@uj.edu.pl}
\thankstext{e-ms}{marek.szydlowski@uj.edu.pl}
\thankstext{e-ab}{andrzej.borowiec@ift.uni.wroc.pl}

\institute{Astronomical Observatory, Jagiellonian University, Orla 171, 30-244 Krakow, Poland \label{oauj}
\and
Mark Kac Complex Systems Research Centre, Jagiellonian University, {\L}ojasiewicza 11, 30-348 Krak{\'o}w, Poland \label{csrc}
\and
Institute for Theoretical Physics, Wroclaw University, pl. Maxa Borna 9, 50-204 Wroclaw, Poland \label{itpwu}}

\date{Received: date / Accepted: date}

\maketitle

\begin{abstract}
We classify singularities in FRW cosmologies, which dynamics can be reduced to the dynamical system of the Newtonian type. This classification is performed in terms of geometry of a potential function if it has poles. At the sewn singularity, which is of a finite scale factor type, the singularity in the past meets the singularity in the future. We show, that such singularities appear in the Starobinsky model in $f(\hat{R})=\hat{R}+\gamma \hat{R}^2$ in the Palatini formalism, when dynamics is determined by the corresponding piece-wise smooth dynamical system. As an effect we obtain a degenerated singularity. Analytical calculations are given for the cosmological model with matter and the cosmological constant. The dynamics of model is also studied using dynamical system methods. From the phase portraits we find generic evolutionary scenarios of the evolution of the Universe. For this model, the best fit value of $\Omega_\gamma=3\gamma H_0^2$ is equal $9.70\times 10^{-11}$. We consider model in both Jordan and Einstein frames. We show that after transition to the Einstein frame we obtain both form of the potential of the scalar field and the decaying Lambda term.
\end{abstract}

\section{Introduction}

The main aim of the paper is the construction of the Starobinsky model with a squared term $\hat{R}^2$ in the Palatini formalism and the investigation of cosmological implications of this model. In this model the inflation phase of evolution of the universe can be obtained by the modification of general relativity in the framework of $f(\hat{R})$ modified gravity theories \cite{DeFelice:2010aj}. In this context, historically the first theory of inflation was proposed by Starobinsky \cite{Starobinsky:1980te}. In the original Starobinsky model the term $R^2 /6 M^2$ was motivated by the conformal anomaly in the quantum gravity. Problem of inflation in $f(R)$ cosmological model are strictly related with the choice of frames. Authors \cite{DeFelice:2010aj} show that CMB spectra in the both Einstein and Jordan frames are different functions of the number of e-foldings until the end of inflation.

Inflation is a hypothesis about the existence of a short
but very fast (of exponential type) accelerated growth of the scale
factor $a(t)$ during the early evolution of the universe, after the Big-Bang but before the radiation-dominated epoch \cite{Guth:1980zm, Linde:1986fd}. It implies $\ddot{a} (t) > 0$.
Irregularities in the early epoch may lead to the formation of structures in the Universe due to the appearance of inflation.

Starobinsky \cite{Starobinsky:1980te} was the first who proposed a very simple theoretical model with one parameter $M$ (energy scale $M$) of such inflation and which is in a good agreement with astronomical data and CMB observation. The Starobinsky model is representing the simplest version of $f(R)$ gravity theories which have been developed considerably in the last decade \cite{Nojiri:2006ri, DeFelice:2010aj, Sotiriou:2008rp}, whose extra term in the Lagrangian is quadratic in the scalar curvature. This model predicts the value of spectral index $n_s= 0.9603 \pm 0.0073$, at the 68\% CL, with deviation from scale-invariance of the primordial power spectrum \cite{Mukhanov:1981xt, Starobinsky:1983zz}

The Starobinsky model is also compatible with Planck 2015 data \cite{Ade:2015rim} and nicely predicts the number $N =50\sim 60$ e-folds between the start and end of inflation \cite{Cheng:2013iya}.

It has been recently investigated some generalization of the Starobinsky inflationary model with a polynomial form of $f(R) = R + \frac{R^2}{6M^{2}} + \frac{\lambda_n}{2n} \frac{R^n}{(3M^2 )^{n-1}}$. It was demonstrated that the slow-roll inflation can be achieved as long as the dimensionless coupling $\lambda_n$ is sufficiently small \cite{Huang:2013hsb}.

The Starobinsky model becomes generic because the smallness of the dimensionless coupling constant $\lambda_n$ does not imply that fine-tuning is necessary \cite{Huang:2013hsb}. The Starobinsky model was developed in many papers \cite{Starobinsky:1983zz, Kofman:1985aw, Ketov:2010qz, Appleby:2009uf, Capozziello:2009hc, Alho:2016gzi, Capozziello:2015wsa}.

In this paper we develop the idea of endogenous inflation as an effect of modification of the FRW equation after the formulation of $f(R)$ cosmological model in the Palatini formalism.

We are looking for an inflation mechanism as a pure dynamical mechanism driven by presence of the additional term (square of the Ricci scalar) in the Lagrangian, without necessity of the choice of a frame (Einstein vs Jordan frame) \cite{Alho:2016gzi, Capozziello:2015wsa, Capozziello:2010sc}.

In the modern cosmology, the most popular trend is to explain the dark energy and the dark matter in terms of some substances, which nature is unknown up to now. Albert Einstein was representing the opposite relational point of view on description of gravity, in which all substantial forms should be eliminated. Such a point of view is called anti-substantialism. Extended $f(R)$ gravity models \cite{Sotiriou:2008rp,Carroll:2004de} offer intrinsic or geometric models of both dark matter and dark energy---the key elements of Standard Cosmological Model. Therefore, the Einstein idea of relational gravity, in which dark matter and dark energy can be interpreted as geometric objects, is naturally realized in $f(\hat{R})$ extended gravity. The methods of dynamical system in the context of investigation dynamics of $f(R)$ models are used since Carroll \cite{Carroll:2004de, Borowiec:2011wd}.

Unfortunately, the metric formulation of extended gravity gives rise to fourth order field equations. To avoid this difficulty, the Palatini formalism can be apply where both the metric $g$ and symmetric connection $\Gamma$ are assumed to be independent dynamical variables. In consequence, one gets a system of second order partial differential equations. The Palatini approach reveals that the early universe inherits properties of the global $\Lambda$CDM evolution.

In the Palatini gravity action for $f(\hat{R})$ gravity is introduced to be
\begin{equation}
S=S_{\text{g}}+S_{\text{m}}=\frac{1}{2}\int \sqrt{-g}f(\hat{R}) d^4 x+S_{\text{m}},\label{action}
\end{equation}
where $\hat{R}=g^{\mu\nu}\hat{R}_{\mu\nu}({\Gamma})$ is the generalized Ricci scalar and $\hat{R}_{\mu\nu}({\Gamma})$ is the Ricci tensor of a torsionless connection $\Gamma$. In this paper, we assume that $8\pi G=c=1$. The equation of motion obtained from the first order Palatini formalism reduces to
\begin{equation}
f'(\hat{R})\hat{R}_{\mu\nu}-\frac{1}{2}f(\hat{R})g_{\mu\nu}=T_{\mu\nu},\label{structural}
\end{equation}
\begin{equation}
\hat{\nabla}_\alpha(\sqrt{-g}f'(\hat{R})g^{\mu\nu})=0,\label{con}
\end{equation}
where $T_{\mu\nu}=-\frac{2}{\sqrt{-g}}\frac{\delta L_{\text{m}}}{\delta g_{\mu\nu}}$ is matter energy momentum tensor, i.e. one assumes that the matter minimally couples to the metric. As a consequence the energy momentum tensor is conserved, i.e.:
$\nabla^\mu T_{\mu\nu}=0$ \cite{Koivisto:2005yk}.
In eq. (\ref{con}) $\hat{\nabla}_\alpha$ means the covariant derivative calculated with respect to $\Gamma$.
In order to solve equation (\ref{con}) it is convenient to introduce a new metric
\begin{equation}
\sqrt{h}h_{\mu\nu}=\sqrt{-g}f'(\hat{R})g_{\mu\nu}
\end{equation}
for which the connection $\Gamma=\Gamma_{L-C}(h)$ is a Levi-Civita connection. As a consequence in dim $M=4$ one gets
\begin{equation}
h_{\mu\nu}=f'(\hat{R})g_{\mu\nu},
\end{equation}
i.e. that both metrics are related by the conformal factor. For this reason one should assume that the conformal factor $f^\prime(\hat R)\neq 0$, so it has strictly positive or negative values.

Taking the trace of (\ref{structural}), we obtain additional so called structural equation
\begin{equation}
f'(\hat{R})\hat{R}-2 f(\hat{R})=T.\label{structural2}
\end{equation}
where $T=g^{\mu\nu}T_{\mu\nu}$.
Because of cosmological applications we assume that the metric $g$ is FRW metric
\begin{equation}\label{frw}
ds^2=-dt^2+a^2(t)\left[\frac{1}{1-kr^2}dr^2+r^2(d\theta^2+\sin^2\theta d\phi^2)\right],
\end{equation}
where $a(t)$ is the scale factor, $k$ is a constant of spatial curvature ($k=0, \pm 1$), $t$ is the cosmological time. For simplicity of presentation we consider the flat model ($k=0$).

As a source of gravity we assume perfect fluid with the energy-momentum tensor
\begin{equation}
T^\mu_\nu=\text{diag}(-\rho,p,p,p),
\end{equation}
where $p=w\rho$, $w=const$ is a form of the equation of state ($w=0$ for dust and $w=1/3$ for radiation).
Formally, effects of the spatial curvature can be also included to the model by introducing a curvature fluid $\rho_{\text{k}}=-\frac{k}{2}a^{-2}$, with the barotropic factor $w=-{1\over 3}$ ($p_{\text{k}}=-\frac{1}{3}\rho_{\text{k}}$). From the conservation condition $T_{\nu;\mu}^{\mu}=0$ we obtain that $\rho=\rho_0 a^{-3(1+w)}$.
Therefore trace $T$ reads as
\begin{equation}
T=\sum_i \rho_{i,0}(3w_i-1)a(t)^{-3(1+w_i)}.
\end{equation}
In what follows we consider visible and dark matter $\rho_{\text{m}}$ in the form of dust $w=0$, dark energy $\rho_\Lambda$ with $w=-1$ and radiation $\rho_{\text{r}}$ with $w=1/3$.

Because a form of the function $f(\hat{R})$ is unknown, one needs to probe it via ensuing cosmological models.
Here we choose the simplest modification of the general relativity Lagrangian
\begin{equation}\label{lag}
f(\hat{R})=\hat{R}+\gamma \hat{R}^2,
\end{equation}
induced by first three terms in the power series decomposition of an arbitrary function $f(R)$.
In fact, since the terms $\hat R^n$ have different physical dimensions, i.e. $[\hat R^n]\neq [\hat R^m]$ for $n\neq m$, one should take instead the function $\hat R_0 f(\hat R/\hat R_0)$ for constructing our Lagrangian, where $\hat{R_0}$ is a constant and $[\hat{R_0}]=[\hat{R}]$. In this case the power series expansion reads: $\hat R_0 f(\hat R/\hat R_0)=\hat R_0\sum_{n=0} \alpha_n (\hat R/\hat R_0)^n=\sum_{n=0} \tilde\alpha_n \hat R^n$, where the coefficients $\alpha_n$ are dimensionless, while $[\tilde\alpha_n]=[\hat R]^{1-n}$ are dimension full.

From the other hand the Lagrangian (\ref{lag}) can be viewed as a simplest deviation, by the quadratic Starobinsky term, from the Lagrangian $\hat{R}$ which provides the standard cosmological model a.k.a. $\Lambda$CDM model. A corresponding solution of the structural equation (\ref{structural2}) 
\begin{equation}\label{sol}
\hat{R}=-T\equiv 4\rho_{\Lambda,0}+\rho_{\text{m},0}a^{-3}.
\end{equation}
is, in fact, exactly the same as for the $\Lambda$CDM model, i.e. with $\gamma=0$. However, the Friedmann equation of the $\Lambda$CDM model (with dust matter, dark energy and radiation)
\begin{equation}\label{lcdm}
H^2=\frac{1}{3}\left(\rho_{\text{r},0}a^{-4}+\rho_{\text{m},0}a^{-3}+\rho_{\Lambda,0}\right)
\end{equation}
is now hardly affected  by the presence of quadratic term. More exactly a counterpart of the above formula in the  model under consideration looks as follows
\begin{multline}
\frac{H^2}{H_0^2}=\frac{b^2}{\left(b+\frac{d}{2}\right)^2}\left[\Omega_{\gamma}(\Omega_{\text{m},0}a^{-3}+\Omega_{\Lambda,0})^2 \right. \\ 
\times \frac{(K-3)(K+1)}{2b}+(\Omega_\text{m,0}a^{-3}+\Omega_{\Lambda,0}) \\
\left. +\frac{\Omega_{\text{r},0}a^{-4}}{b}+
\Omega_k\right],\label{friedmann2}
\end{multline}
where
\begin{align}
\Omega_k &= -\frac{k}{H_0^2 a^2},\\
\Omega_{\text{r},0} &= \frac{\rho_\text{r,0}}{3H_0^2},\\
\Omega_{\text{m},0} &= \frac{\rho_\text{m,0}}{3H_0^2},\\
\Omega_{\Lambda,0} &= \frac{\rho_{\Lambda,0}}{3H_0^2},\\
K &= \frac{3\Omega_{\Lambda,0}}{(\Omega_\text{m,0}a^{-3}+\Omega_{\Lambda,0})},\\
\Omega_{\gamma} &= 3\gamma H_0^2,\\
b &= f'(\hat{R})=1+2\Omega_\gamma(\Omega_\text{m,0}a^{-3}+4\Omega_{\Lambda,0}),\\
d &= \frac{1}{H}\frac{db}{dt}=-2\Omega_{\gamma}(\Omega_\text{m,0}a^{-3}+\Omega_{\Lambda,0})(3-K)
\end{align}
From the above one can check that the standard model (\ref{lcdm}) can be reconstructed in the limit $\gamma\mapsto 0$. The study of this generalized Friedmann equation is a main subject of our research.

The paper is organized as follows. In section 2, we consider the Palatini approach in the Jordan and Einstein frame. In section 3, we present some generalities concerning dynamical systems of Newtonian type, and their relations with the Palatini-Starobinsky model. Section 4, is devoted to classification of cosmological singularities with special attention on Newtonian type systems represented by potential function $V(a)$. We adopt the Fernandes-Jambrina and Lazkoz classification of singularities \cite{FernandezJambrina:2006hj} to these systems using the notion of elasticity of the potential function with respect the scale factor. In section 5, we will analyze the singularities in the Starobinsky model in the Palatini formalism. This system requires the form of piece-wise smooth dynamical system. Statistical analysis of the model is presented in section 6. In section 7, we shall summarize obtained results and draw some conclusions.

\section{The Palatini approach in different frames (Jordan vs Einstein frame)}

Because the effect of acceleration can depend on a choice of a frame \cite{Bahamonde:2017kbs} this section is devoted to showing the existence of the inflation effect if the model is considered in the Einstein frame.

The action (\ref{action}) is dynamically equivalent to the first order Palatini gravitational action, provided that $f^{''}(\hat R) \neq 0 $ \cite{DeFelice:2010aj, Sotiriou:2008rp, Capozziello:2015wsa}
\begin{multline}\label{action1}
 S(g_{\mu\nu}, \Gamma^\lambda_{\rho\sigma}, \chi)=\frac{1}{2}\int\mathrm{d}^4x\sqrt{-g}\left(f^\prime(\chi)(\hat R-\chi) + f(\chi) \right) \\ + S_m(g_{\mu\nu},\psi),
\end{multline}
Introducing a scalar field $\Phi=f'(\chi)$ and taking into account the constraint $\chi=\hat R$, one gets the action (\ref{action1}) in the following form
\begin{multline}\label{actionP}
 S(g_{\mu\nu}, \Gamma^\lambda_{\rho\sigma},\Phi)=\frac{1}{2}\int\mathrm{d}^4x\sqrt{-g}\left(\Phi \hat R - U(\Phi) \right) \\ + S_m(g_{\mu\nu},\psi),
\end{multline}
where the potential $U(\Phi)$ is defined by
\begin{equation}\label{PotentialP}
 U_f(\Phi)\equiv U(\Phi)=\chi(\Phi)\Phi-f(\chi(\Phi))
\end{equation}
with $\Phi = \frac{d f(\chi)}{d\chi}$ and $\hat R\equiv \chi = \frac{d U(\Phi)}{d\Phi}$.

The Palatini variation of the action (\ref{actionP}) gives rise to the following equations of motion
\begin{subequations}	
 \begin{align}
	\label{EOM_P}
	\Phi\left( \hat R_{\mu\nu} - \frac{1}{2} g_{\mu\nu} \hat R \right) &  +{1\over 2} g_{\mu\nu} U(\Phi) - T_{\mu\nu} = 0,\\
	\label{EOM_connectP}
	& \hat{\nabla}_\lambda(\sqrt{-g}\Phi g^{\mu\nu})=0,\\
	%
	\label{EOM_scalar_field_P}
	  \hat R &  -  U^\prime(\Phi) =0.
	\end{align}
\end{subequations}
Equation (\ref{EOM_connectP}) implies that the connection $\hat \Gamma$ is a metric connection for a new metric $\bar g_{\mu\nu}=\Phi g_{\mu\nu}$; thus $\hat R_{\mu\nu}=\bar R_{\mu\nu}, \bar R= \bar g^{\mu\nu}\bar R_{\mu\nu}=\Phi^{-1} \hat R$ and $\bar g_{\mu\nu}\bar R=\ g_{\mu\nu}\hat R$.
The $g$-trace of (\ref{EOM_P}) produces a new structural equation
\begin{equation}\label{struc2}
  2U(\Phi)-U'(\Phi)\Phi=T.
\end{equation}
Now equations (\ref{EOM_P}) and (\ref{EOM_scalar_field_P}) take the following form
	\begin{align}
	\label{EOM_P1}
	 \bar R_{\mu\nu} - \frac{1}{2} \bar g_{\mu\nu} \bar R  &  = \bar T_{\mu\nu}-{1\over 2} \bar g_{\mu\nu} \bar U(\Phi),\\
 	%
	\label{EOM_scalar_field_P1}
	  \Phi\bar R &  -  (\Phi^2\,\bar U(\Phi))^\prime =0,
	\end{align}
where we introduce $\bar U(\phi)=U(\phi)/\Phi^2$, $\bar T_{\mu\nu}=\Phi^{-1}T_{\mu\nu}$ and the structural equation can be replaced by
\begin{equation}\label{EOM_P1c}
 \Phi\,\bar U^\prime(\Phi)  + \bar T = 0\,.
\end{equation}
In this case, the action for the metric $\bar g_{\mu\nu }$ and scalar field $\Phi$ is given by
\begin{equation}\label{action2}
 S(\bar g_{\mu\nu},\Phi)=\frac{1}{2}\int\mathrm{d}^4x\sqrt{-\bar g}\left(\bar R- \bar U(\Phi) \right) + S_m(\Phi^{-1}\bar g_{\mu\nu},\psi),
\end{equation}
where we have to take into account a non-minimal coupling between $\Phi$ and $\bar g_{\mu\nu}$
\begin{equation}\label{em_2}
    \bar T^{\mu\nu} =
-\frac{2}{\sqrt{-\bar g}} \frac{\delta}{\delta \bar g_{\mu\nu}}S_m  = (\bar\rho+\bar p)\bar u^{\mu}\bar u^{\nu}+ \bar p\bar g^{\mu\nu}=\Phi^{-3}T^{\mu\nu}~,
\end{equation}
$\bar u^\mu=\Phi^{-{1\over 2}}u^\mu$, $\bar\rho=\Phi^{-2}\rho,\ \bar p=\Phi^{-2}p$, $\bar T_{\mu\nu}= \Phi^{-1}T_{\mu\nu}, \ \bar T= \Phi^{-2} T$ (see e.g. \cite{Capozziello:2015wsa, Dabrowski:2008kx}).

In FRW case, one gets the metric $\bar g_{\mu\nu}$ in the following form
\begin{equation}\label{frwb}
d\bar s^2=-d\bar t^2+\bar a^2(t)\left[dr^2+r^2(d\theta^2+\sin^2\theta d\phi^2)\right],
\end{equation}
where $d\bar t=\Phi(t)^{1\over 2}d\, t$ and new scale factor $\bar a(\bar t)=\Phi(\bar t)^{1\over 2}a(\bar t) $. 	
Ensuing cosmological equations (in the case of the barotropic matter) are given by
\begin{equation}\label{frwb2}
3\bar H^2= \bar \rho_\Phi + \bar\rho _m, \quad 6\frac{\ddot{\bar a}}{\bar a}=2\bar\rho_\Phi -\bar{\rho}_m(1+3w)
\end{equation}
where
\begin{equation}\label{frwb3}
\bar\rho_\Phi={1\over 2}\bar U(\Phi),\quad \bar{\rho}_{\text{m}}=\rho_0\bar a^{-3(1+w)}\Phi^{{1\over 2}(3w-1)}\end{equation}
and
$w=\bar p_{\text{m}} / \bar\rho_{\text{m}}= p_{\text{m}} / \rho_{\text{m}}$. In this case, the conservation equations has the following form
\begin{equation}\label{frwb4}
\dot{\bar{\rho}}_{\text{m}}+3\bar H\bar{\rho}_{\text{m}}(1+w)=-\dot{\bar{\rho}}_\Phi.\end{equation}

\begin{figure}
	\centering
	\includegraphics[width=0.7\linewidth]{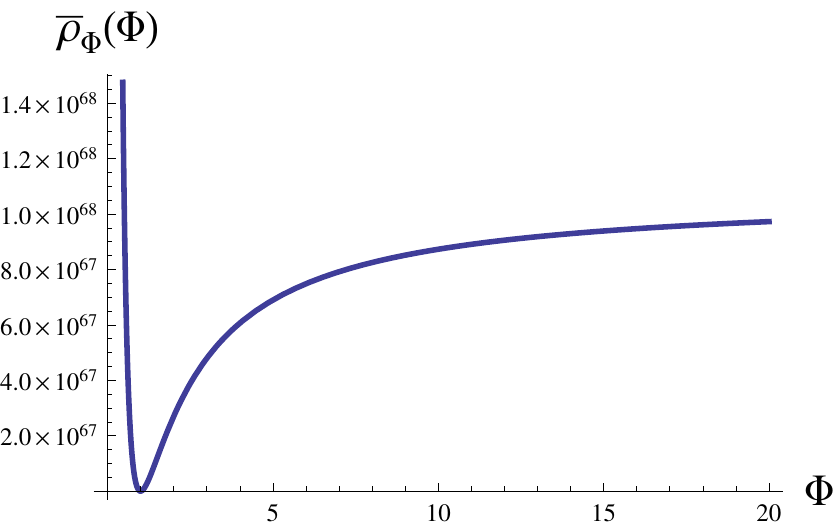}
	\caption{Illustration of the dependence $\bar\rho_\Phi$ of $\Phi$. We assume that $\gamma=1.16\times 10^{-69}\text{ s}^2$. The units of $\bar\rho_\Phi$ are expressed in $\frac{\text{km}^2}{\text{s}^2 \text{Mpc}^2}$. Note that this potential has the same shape like the Starobinsky potential.}
	\label{fig:16}
\end{figure}

Let us consider the Starobinsky--Palatini model in the above framework. The potential $\bar U$ is described by the following formula
\begin{equation}
\bar U(\Phi)=2\bar\rho_\Phi(\Phi)=\left(\frac{1}{4\gamma}+2\lambda\right)\frac{1}{\Phi^2}-\frac{1}{2\gamma}\frac{1}{\Phi}+
\frac{1}{4\gamma}.
\end{equation}
Figure \ref{fig:16} presents the relation $\bar\rho_\Phi(\Phi)$. Note that the function $\bar\rho_\Phi$ has the same shape like the Starobinsky potential. The  function $\bar\rho_\Phi(\Phi)$ has the minimum for
\begin{equation}
\Phi_{\text{min}}=1+8\gamma\lambda.
\end{equation}
In general, the scalar field $\Phi(\bar a)$ is given by (cf. (\ref{sol}))
\begin{equation} \label{eq:2.14}
\Phi=1+2\gamma\hat R= 1+8\gamma\lambda+2\gamma \rho_m -6\gamma p_m.
\end{equation}
Because $\bar\rho_m=\Phi^{-2}\rho_m,\ \bar p_m=\Phi^{-2}p_m$, and taking into  account (\ref{frwb3}) one gets 
\begin{equation} \label{eq:2.15}
2\gamma(1-3w)\rho_0\bar a^{-3(1+w)}\Phi^{{3\over 2}(w+1)}-\Phi+1+8\gamma\lambda=0.
\end{equation}
the algebraic equation determining the function $\Phi(\bar a)$ for a given barotropic factor $w$.
This provides an implicit dependence $\Phi(\bar a)$. In order to get it more explicit one needs to solve (\ref{eq:2.15}) for some interesting values $w$. For example in the case of dust we obtain the third order polynomial equation
$$({1\over 2\gamma} +4\Lambda)y^3-{1\over 2\gamma}y+\rho_{0w}\bar a^{-3}=0$$
where $y=\Phi^{-{1\over 2}}$.

\begin{figure}
	\centering
	\includegraphics[width=0.7\linewidth]{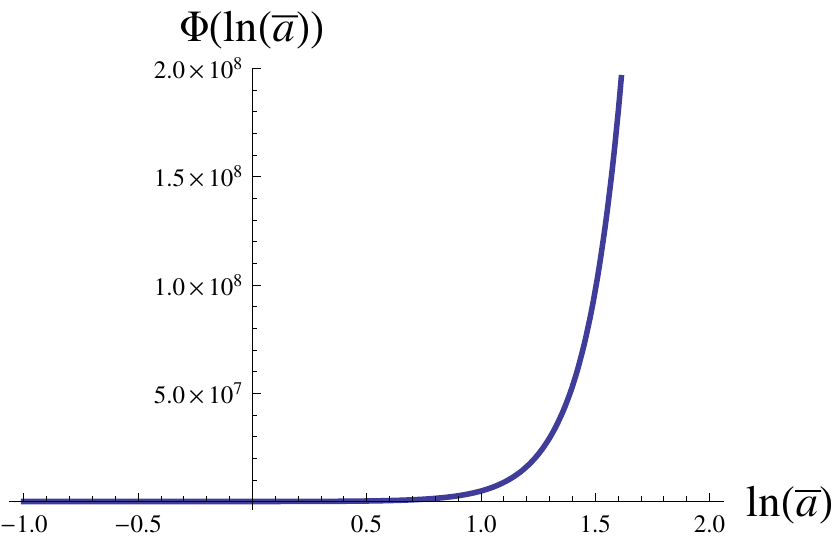}
	\caption{Illustration of the typical evolution of $\Phi$ with respect to $\ln(\bar{a})$ at the beginning of the inflation epoch. We assume that $\gamma=1.16\times 10^{-69}\text{ s}^2$ and $\bar a_0=1$ at the beginning of the inflation epoch.}
	\label{fig:12}
\end{figure}

The evolution of $\Phi(\bar a)$, at the beginning of the inflation epoch, is presented in figure \ref{fig:12}.


For $\gamma\approx 0$, the potential $\bar{U}$ can be approximated as $\bar{U}=-\bar{\rho}_m+\frac{1}{4\gamma}$.
In this case the Friedmann equation can be written as
\begin{equation}
3\bar{H}^2=\frac{\bar{\rho}_m}{2}+\frac{1}{8\gamma}.
\end{equation}

In the case of $\bar\rho_m=0$, $\bar\rho_\Phi$ is constant and the Friedmann equation has the following form
\begin{equation}
3\bar H^2=\frac{1}{8\gamma}.
\end{equation}

\begin{figure}
	\centering
	\includegraphics[width=0.7\linewidth]{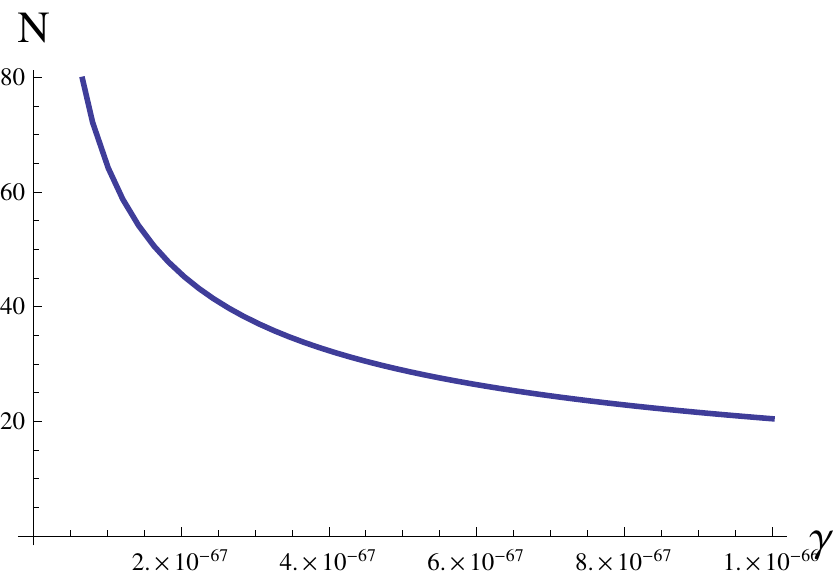}
	\caption{The diagram of the relation between $\gamma$ and the approximate number of e-foldings $N=\bar H_{\text{init}} (\bar t_{\text{fin}} -\bar t_{\text{init}})$ from $\bar t_\text{init}$ to $\bar t_\text{fin}$. We assume that $\bar t_{\text{fin}} -\bar t_{\text{init}}\approx10^{-32}\text{ s}$. The units of the parameter $\gamma$ are expressed in $\text{s}^2$. Note that the number of e-foldings grows when the parameter $\gamma$ decreases and $N=60$ when $\gamma=1.16\times 10^{-69}\text{ s}^2$.}
	\label{fig:17}
\end{figure}

In this model the inflation phenomenon appears when the the value of the parameter $\gamma$ is close to zero and the matter $\bar{\rho}_m$ is negligible with comparison to $\bar\rho_\Phi$. In this case the approximate number of e-foldings is given by the following formula
\begin{equation}
N=H_\text{init}(\bar t_\text{fin}-\bar t_\text{init})=\frac{\bar t_\text{fin}-\bar t_\text{init}}{\sqrt{24\gamma}}.
\end{equation}
The number of e-folds $N$ should be equal $50\sim 60$ in the inflation epoch \cite{Cheng:2013iya}. In this model we obtain $N=60$, when $\gamma=1.16\times 10^{-69}\text{ s}^2$ and the timescale of the inflation  is equal $10^{-32}\text{ s}$ \cite{Ade:2014xna}. The relation between $\gamma$ and the approximate number of e-foldings $N$ is presented in figure \ref{fig:17}.

The condition for appearing of the inflation requires the value of the parameter $\gamma$ to be close to zero, hence the influence of the parameter $\lambda$ for the evolution of the universe is negligible.

\begin{figure}
	\centering
	\includegraphics[width=0.7\linewidth]{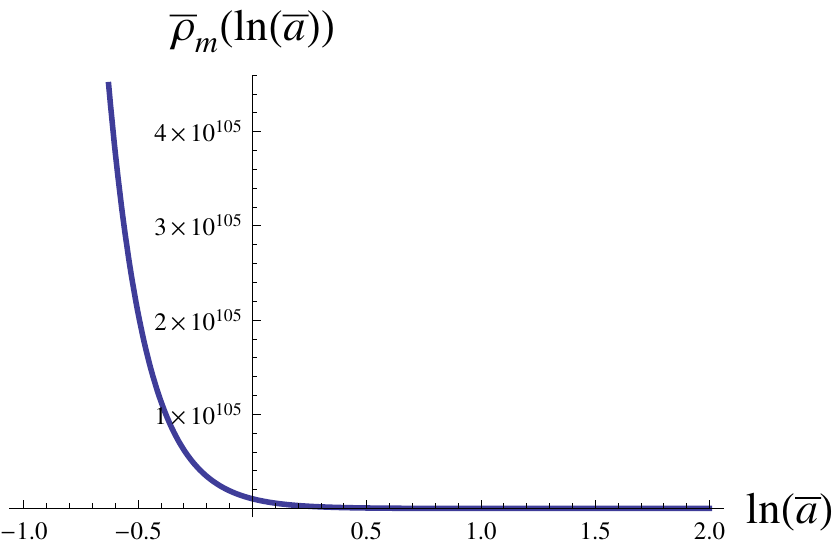}
	\caption{Illustration of the typical evolution of $\bar\rho_m$ with respect to $\ln(\bar{a})$ at the beginning of the inflation epoch. We assume that $\gamma=1.16\times 10^{-69}\text{ s}^2$ and $\bar a_0=1$ at the beginning of the inflation epoch. The units of $\bar\rho_m$ are expressed in $\frac{\text{km}^2}{\text{s}^2 \text{Mpc}^2}$.}
	\label{fig:11}
\end{figure}

\begin{figure}
	\centering
	\includegraphics[width=0.7\linewidth]{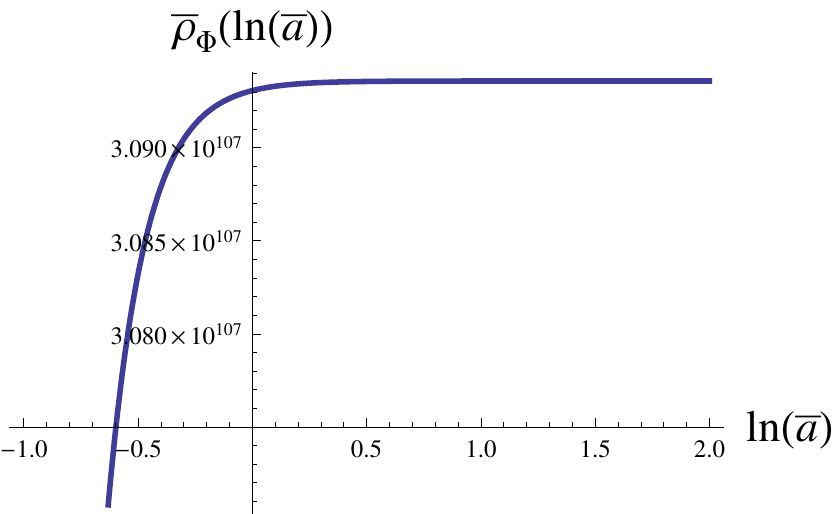}
	\caption{Illustration of the typical evolution of $\bar\rho_\phi$ with respect to $\ln(\bar{a})$ at the beginning of the inflation epoch. We assume that $\gamma=1.16\times 10^{-69}\text{ s}^2$ and $\bar a_0=1$ at the beginning of the inflation epoch. The units of $\bar\rho_\Phi$ are expressed in $\frac{\text{km}^2}{\text{s}^2 \text{Mpc}^2}$. Note that during the inflation $\bar\rho_\phi\approx {\text{const}}$.}
	\label{fig:13}
\end{figure}

\begin{figure}
	\centering
	\includegraphics[width=0.7\linewidth]{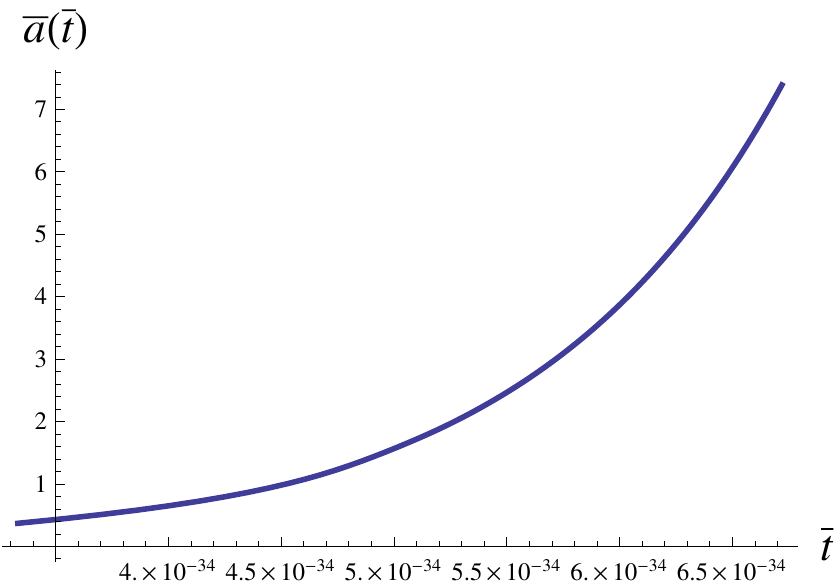}
	\caption{Illustration of the typical evolution of $\bar a$ with respect to $\bar{t}$ at the beginning of the inflation epoch. We assume that $\gamma=1.16\times 10^{-69}\text{ s}^2$ and $\bar a_0=1$ at the beginning of the inflation epoch. The time $\bar t$ is expressed in seconds.}
	\label{fig:15}
\end{figure}

\begin{figure}
	\centering
	\includegraphics[width=0.7\linewidth]{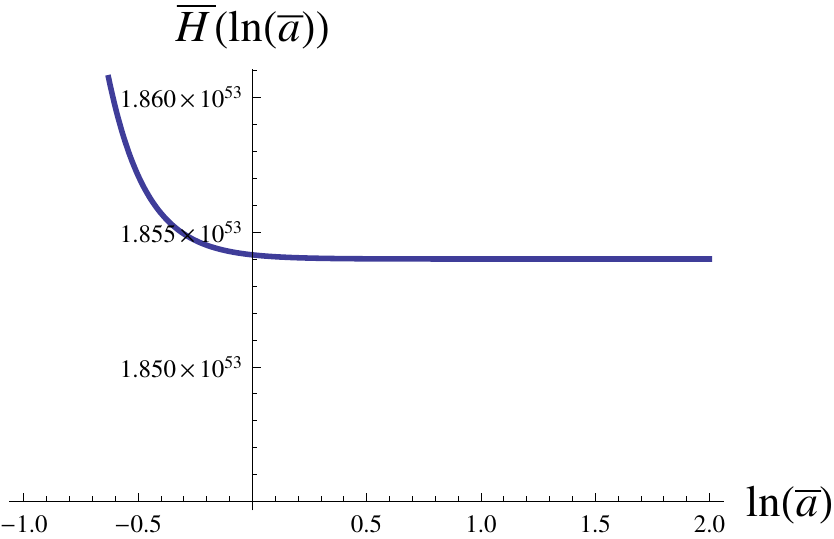}
	\caption{Illustration of the typical evolution of $\bar H$ with respect to $\ln(\bar{a})$ at the beginning of the inflation epoch. We assume that $\gamma=1.16\times 10^{-69}\text{ s}^2$ and $\bar a_0=1$ at the beginning of the inflation epoch. The units of $\bar H$ are expressed in $\frac{\text{km}}{\text{s } \text{Mpc}}$. Note that for the late time, $\bar H$ can be treated as a constant.}
	\label{fig:14}
\end{figure}

In figure \ref{fig:11} it is demonstrated the typical evolution of $\bar\rho_m(\bar a)$ at the beginning of the inflation epoch. The typical evolution of $\bar \rho_\Phi$, at the beginning of the inflation epoch, is presented in figure \ref{fig:13}. Note that, for the late time universe $\bar \rho_\Phi$ can be approximated as a constant. Figure \ref{fig:15} presents the evolution of the scale factor $\bar{a}(\bar t)$ during the inflation. Figure \ref{fig:14} shows the Hubble function $\bar{H}$ during the inflation epoch.

\begin{figure}
	\centering
	\includegraphics[width=0.7\linewidth]{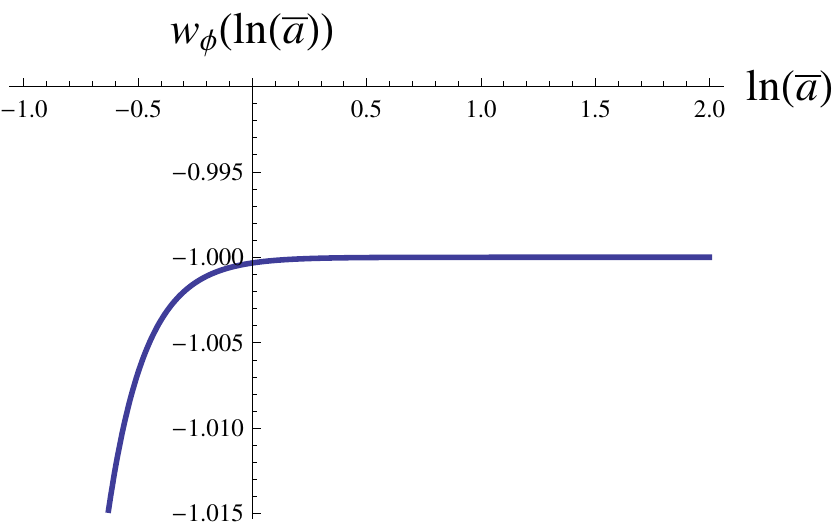}
	\caption{Illustration of the typical evolution of $w_\phi$ with respect to $\ln(\bar{a})$. We assume that $\gamma=1.16\times 10^{-69}\text{ s}^2$ and $\bar a_0=1$ at the beginning of the inflation epoch. Note that during the inflation $w_\phi\approx -1$.}
	\label{fig:18}
\end{figure}

The conservation equation for $\bar\rho_\Phi$ can be written as
\begin{equation}
\dot{\bar\rho}_\Phi=-3\bar H (\bar\rho_\Phi+\bar p_\Phi),
\end{equation}
where $\bar p_\Phi$ is an effective pressure. In this case the equation of state for the dark energy is expressed by the following formula
\begin{equation}
\bar p_\Phi=w(a)\bar\rho_\Phi,
\end{equation}
where the function $w(a)$ is given by the expression
\begin{equation}
w(a)=-1-\frac{\dot{\bar\rho}_\Phi}{\sqrt{3}\sqrt{\bar\rho_\text{m}+\bar\rho_\Phi}\rho_\Phi}=-1-\frac{1}{3\bar H}\frac{d\ln\rho_\Phi}{d\bar t}.
\end{equation}
The diagram of coefficient of equation of state $w(a)$, at the beginning the inflation epoch, is presented in figure \ref{fig:18}. Note that the function $w(a)$, for the late time, is a constant and equal $-1$.

The action (\ref{actionP}) can be rewritten in the Jordan frame $(g_{\mu\nu }, \Phi)$ as
\begin{equation}\label{action3}
 S=\frac{1}{2k}\int\mathrm{d}^4x\sqrt{-g}\left(\Phi R +\frac{3}{2\Phi}\partial_\mu\Phi\partial^\mu\Phi-
 U(\Phi) \right),
\end{equation}
where $R$ is the metric Ricci scalar, $\Phi=f'(\hat{R})$, $\hat{R}=\chi(\Phi)$.

We obtain Brans-Dicke action with the coupling parameter $\omega=-\frac{3}{2}$ in the Jordan frame. Equations of motion take the form
\begin{subequations}
	\label{EOM}
	\begin{multline}
	\Phi\left( R_{\mu\nu} - \frac{1}{2} g_{\mu\nu} R \right) - \frac{3}{4\Phi}g_{\mu\nu} \nabla_\sigma \Phi \nabla^\sigma \Phi +\frac{3}{2\Phi}\nabla_\mu \Phi \nabla_\nu \Phi \\
	\label{EOM_metric}
	+  g_{\mu\nu}\Box \Phi - \nabla_\mu\nabla_\nu\Phi +{1\over 2} g_{\mu\nu} U(\phi) - \kappa T_{\mu\nu} = 0 \,,
	\end{multline}
	\begin{equation}
	\label{EOM_scalar_field}
	  R -\frac{3}{\Phi}\Box \Phi +  \frac{3}{2\Phi^2} \nabla_\mu \Phi \nabla^\mu \Phi - {1\over 2} U^\prime(\Phi) =0. 
	\end{equation}
\end{subequations}

In this case the dynamics of the metric $g$ is exactly the same as described by the original Palatini equations (\ref{structural}) -- (\ref{structural2}).
On the cosmological ground it means that the scale factor $a(t)$ evolves according to  the Friedmann equation (\ref{friedmann2}). It has been recently shown that cosmological data favor the value $\omega\approx -1$ on the $3\sigma$ level \cite{Hrycyna:2014cka}.

\section{Singularities in cosmological dynamical systems of Newtonian type}

There is a class of cosmological models, which dynamics can be reduced to the dynamical system of the Newtonian type. Let consider a homogeneous and isotropic universe with a spatially flat space-time metric of the form
\begin{equation}\label{frw2}
ds^2=dt^2-a^2(t)\left[dr^2+r^2(d\theta^2+\sin^2\theta d\phi^2)\right],
\end{equation}
where $a(t)$ is the scale factor and $t$ is the cosmological time.

Let us consider the energy-momentum tensor $T_\nu^\mu$ for the perfect fluid with energy density $\rho(t)$ and pressure $p(t)$ as a source of gravity. In this case the Einstein equations assumes the form of Friedmann equations
\begin{align}
\rho &= 3H^2=\frac{3\dot{a}^2}{a^2},\label{hubble} \\
p &= -\frac{2\ddot{a}}{a}-\frac{\dot{a}^2}{a^2},\label{pressure}
\end{align}
where dot denotes differentiation with respect to the cosmic time $t$, $H\equiv\frac{\dot{a}}{a}$ is the Hubble function.
In our notation we use the natural system of units in which $8\pi G=c=1$.

We assume that $\rho(t)=\rho(a(t))$ as well as $p(t)=p(a(t))$, i.e. that both energy density as well as pressure depends on the cosmic time through the scale factor $a(t)$.
The conservation condition $T^{\mu\nu}_{;\mu}=0$ reduces to
\begin{equation}
\dot{\rho}=-3H(\rho+p).\label{cons}
\end{equation}

It would be convenient to rewrite (\ref{hubble}) in an equivalent form
\begin{equation}
\dot{a}^2=-2V(a),\label{friedmann}
\end{equation}
where
\begin{equation}
V(a)=-\frac{\rho(a)a^2}{6}.\label{potential}
\end{equation}
In (\ref{potential}) $\rho(a)$ plays the model role of effective energy density. For example for the standard cosmological model (\ref{lcdm})
\begin{equation}
V=-\frac{\rho_{\text{eff}} a^2}{6}=-\frac{a^2}{6}\left(\rho_{m,0}a^{-3}+\rho_{\Lambda,0}\right),
\end{equation}
where $\rho_{\text{eff}}=\rho_{\text{m}}+\Lambda$ and $\rho_{\text{m}}=\rho_{\text{m,0}}a^{-3}$.
Equation (\ref{pressure}) is equivalent to
\begin{equation}
\frac{\ddot{a}}{a}=-\frac{1}{6}(\rho+3p),
\end{equation}
which is called acceleration equation. It is easily to check that
\begin{equation}
\ddot{a}=-\frac{\partial V}{\partial a},\label{pot}
\end{equation}
where $V(a)$ is given by (\ref{potential}) provided that conservation equation (\ref{cons}) is fulfilled.

Due to equation (\ref{pot}) the evolution of a universe can be interpreted as a motion of a fictitious particle of unit mass in the potential $V(a)$. Here $a(t)$ plays the role of a positional variable. Equation of motion (\ref{pot}) assumes the form analogous to the Newtonian equation of motion.

If we know the form of effective energy density then we can construct the form of potential $V(a)$, which determine the whole dynamics in the phase space $(a, \dot{a})$. In this space the Friedmann equation (\ref{friedmann}) plays the role of a first integral and determines the phase space curves representing the evolutionary paths of the cosmological models. The diagram of potential $V(a)$ contains all information needed to construction of a phase space portrait.
In this case the phase space is two-dimensional
\begin{equation}
	\left\{ (a,\dot{a}) \colon \frac{\dot{a}^2}{2}+V(a)=-\frac{k}{2} \right\}.
\end{equation}
In a general case of arbitrary potential, a dynamical system which describes the evolution of a universe takes the form
\begin{align}
\dot{a} &= x,\label{dota} \\
\dot{x} &= -\frac{\partial V(a)}{\partial a} .\label{dotx}
\end{align}
We shall study the system above using theory of piece-wise smooth dynamical systems. Therefore it is assumed that the potential function, except some isolated (singular) points, belongs to the class $C^2(\mathbb{R}_+)$.

The lines $\frac{x^2}{2}+V(a)=-\frac{k}{2}$ represent possible evolutions of the universe for different initial conditions. The equations (\ref{dota}) and (\ref{dotx}) can be rewritten in dimensionless variables if we replace effective energy density $\rho_\text{eff}$ by density parameter
\begin{equation}
\Omega_\text{eff}=\frac{\rho_\text{eff}}{3H_0^2},
\end{equation}
then
\begin{align}
\frac{1}{H_0^2}\frac{\dot{a}^2}{2} &= -\frac{\Omega_\text{eff} a^2}{2},\\
\frac{d^2 a}{d\tau^2} &= -\frac{\partial \tilde{V}}{\partial a},
\end{align}
where $t\rightarrow\tau=|H_0|t$ and
\begin{equation}
\tilde{V}(a)=-\frac{\Omega_\text{eff} a^2}{2}.
\end{equation}

Any cosmological model can be identified by its form of the potential function $V(a)$ depending on the scale factor $a$. From the Newtonian form of the dynamical system (\ref{dota})-(\ref{dotx}) one can see that all critical points correspond to vanishing of r.h.s of the dynamical system $\left(x_0=0,\text{ }\frac{\partial V(a)}{\partial a}|_{a=a_0}\right)$. Therefore all critical points are localized on the $x$-axis, i.e. they represent a static universe.

Because of the Newtonian form of the dynamical system the character of critical points is determined from the characteristic equation of the form
\begin{equation}
a^2+\det A|_{x_0=0,\frac{\partial V(a)}{\partial a}|_{a_0}=0}=0,\label{det}
\end{equation}
where $\det A$ is determinant of linearization matrix calculated at the critical points, i.e.
\begin{equation}
\det A=\frac{\partial^2 V(a)}{\partial a^2}|_{a_0,\frac{\partial V(a)}{\partial a}|_{a_0}=0}.\label{det2}
\end{equation}
From equation (\ref{det}) and (\ref{det2}) one can conclude that only admissible critical points are the saddle type if $\frac{\partial^2 V(a)}{\partial a^2}|_{a=a_0}<0$ or centers type if $\frac{\partial^2 V(a)}{\partial a^2}|_{a=a_0}>0$.

If a shape of the potential function is known (from the knowledge of effective energy density), then it is possible to calculate cosmological functions in exact form
\begin{equation}
t=\int^a \frac{da}{\sqrt{-2V(a)}},
\end{equation}
\begin{equation}
H(a)=\pm\sqrt{-\frac{2V(a)}{a^2}},
\end{equation}
a deceleration parameter, an effective barotropic factor
\begin{equation}
q=-\frac{a\ddot{a}}{\dot{a}^2}=\frac{1}{2}\frac{d\ln(-V)}{d\ln a},
\end{equation}
\begin{equation}
w_\text{eff}(a(t))=\frac{p_\text{eff}}{\rho_{\text{eff}}}=-\frac{1}{3}\left(\frac{d\ln(-V)}{d\ln a}+1\right),
\end{equation}
a parameter of deviation from de Sitter universe \cite{FernandezJambrina:2006hj}
\begin{equation}
h(t)\equiv-(q(t)+1)=\frac{1}{2}\frac{d\ln(-V)}{d\ln a} - 1
\end{equation}
(note that if $V(a)=-\frac{\Lambda a^2}{6}$, $h(t)=0$), effective matter density and pressure
\begin{equation}
\rho_{\text{eff}}=-\frac{6 V(a)}{a^2},
\end{equation}
\begin{equation}
p_{\text{eff}}=\frac{2 V(a)}{a^2}\left(\frac{d\ln(-V)}{d\ln a}+1\right)
\end{equation} and, finally, a Ricci scalar curvature for the FRW metric (\ref{frw2})
\begin{equation}
R=\frac{6 V(a)}{a^2}\left(\frac{d\ln(-V)}{d\ln a}+2\right).
\end{equation}
	
From the formulas above one can observe that the most of them depend on the quantity
\begin{equation}
I_\nu(a)=\frac{d\ln(-V)}{d\ln a}.
\end{equation}
This quantity measures elasticity of the potential function, i.e. indicates how the potential $V(a)$ changes if the scale factor $a$ changes. For example, for the de Sitter universe $-V(a)\propto a^2$, the rate of growth of the potential is $2\%$ as the rate of growth of the scale factor is $1\%$.

In the classification of the cosmological singularities by Fernandez-Jambrina and Lazkoz \cite{FernandezJambrina:2006hj} the crucial role is played by the parameter $h(t)$. Note that a cosmological sense of this parameter is
\begin{equation}
h(t)=\frac{1}{2}I_\nu (a) -1.
\end{equation}
In this approach the classification of singularities is based on generalized power and asymptotic expansion of the barotropic index $w$ in the equation of state (or equivalently of the deceleration parameter $q$) in terms of the time coordinate.

\section{Degenerated singularities -- new type (VI) of singularity -- sewn singularities}

Recently, due to discovery of an accelerated phase in the expansion of our Universe, many theoretical possibilities for future singularity are seriously considered. If we assume that the Universe expands following the Friedmann equation, then this phase of expansion is driven by dark energy -- hypothetical fluid, which violates the strong energy condition. Many of new types of singularities were classified by Nojiri et al. \cite{Nojiri:2005sx}. Following their classification the type of singularity depends on the singular behavior of the cosmological quantities like: the scale factor $a$, the Hubble parameter $H$, the pressure $p$ and the energy density $\rho$:
\begin{itemize}
	\item Type 0: `Big crunch'. In this type, the scale factor $a$ is vanishing and blow up of the Hubble parameter $H$, energy density $\rho$ and pressure $p$.
	\item Type I: `Big rip'. In this type, the scale factor $a$, energy density $\rho$ and pressure $p$ are blown up.
	\item Type II: `Sudden'. The scale factor $a$, energy density $\rho$ and Hubble parameter $H$ are finite and $\dot{H}$ and the pressure $p$ are divergent.
	\item Type III: `Big freeze'. The scale factor $a$ is finite and the Hubble parameter $H$, energy density $\rho$ and pressure $p$ are blown up \cite{Barrow:2004xh} or divergent \cite{BouhmadiLopez:2006fu}.
	\item Type IV. The scale factor $a$, Hubble parameter $H$, energy density $\rho$, pressure $p$ and $\dot{H}$ are finite but higher derivatives of the scale factor $a$ diverge.
	\item Type V. The scale factor $a$ is finite but the energy density $\rho$ and pressure $p$ vanish.
\end{itemize}

\begin{figure}
	\centering
	\includegraphics[width=0.7\linewidth]{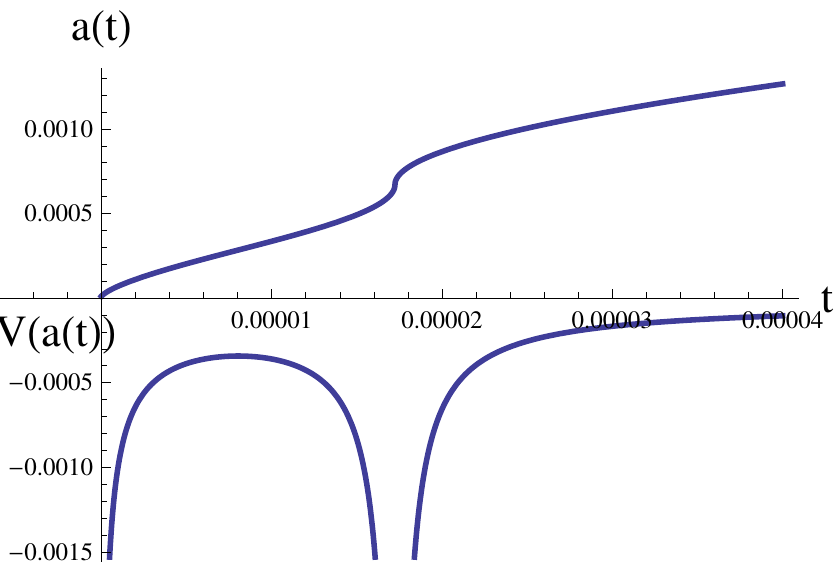}
	\caption{Illustration of sewn freeze singularity, when the potential $V(a)$ has a pole.}
	\label{fig:1}
\end{figure}

\begin{figure}
	\centering
	\includegraphics[width=0.7\linewidth]{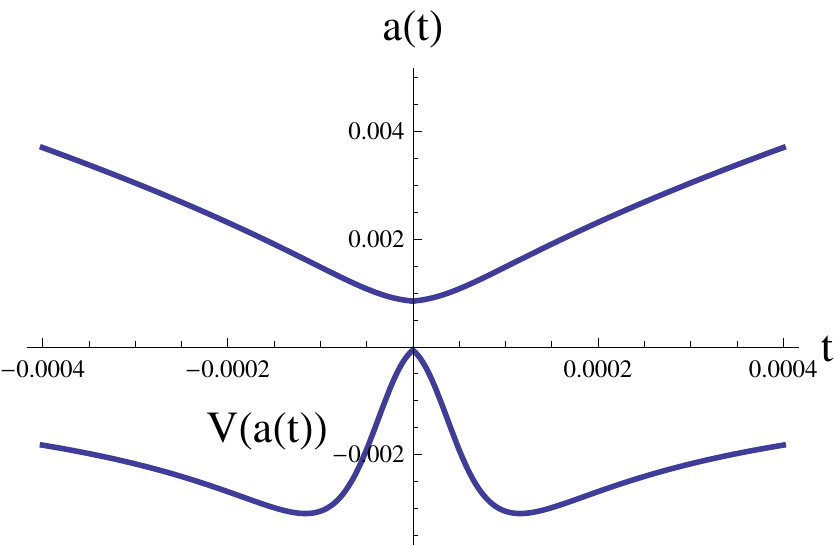}
	\caption{Illustration of a sewn sudden singularity. The model with negative $\Omega_\gamma$ has a mirror symmetry with respect to the cosmological time. Note that the spike on the diagram shows discontinuity of the function $\frac{\partial V}{\partial a}$. Note the existence of a bounce at $t=0$.}
	\label{fig:6}
\end{figure}

Following Kr{\'o}lak \cite{Krolak:1986tp}, big rip and big crunch singularities are strong whereas sudden, big freeze and type IV are weak singularities.

In the model under consideration the potential function or/and its derivative can diverge at isolated points (value of the scale factor). Therefore mentioned before classification has application only for a single component of piece-wise smooth potential. In our model the dynamical system describing evolution of a universe belongs to the class of a piecewise smooth dynamical systems. As a consequence new types of singularities at finite scale factor $a_s$ can appear for which $\frac{\partial V}{\partial a}(a_s)$ does not exist (is not determined). This implies that the classification of singularities should be extended to the case of non-isolated singularities.

Let us illustrate this idea on the example of freeze singularity in the Starobinsky model with the Palatini formalism (previous section). Such a singularity has a complex character and in analogy to the critical point we called it degenerated. Formally it is composed of two types III singularities: one in the future and another one in the past. If we considered the evolution of the universe before this singularity we detect isolated singularity of type III in the future. Conversely if we consider the evolution after the singularity then going back in time we meet type III singularity in the past. Finally, at the finite scale factor both singularities will meet together.
For description of behavior near the singularity one considers $t=t(a)$ relation. This relation has a horizontal inflection point and it is natural to expand this relation in the Taylor series near this point at which $\frac{dt}{da}=\frac{1}{Ha}$ is zero. For the freeze singularity, the scale factor remains constant $a_\text{s}$, $\rho$ and $H$ blow up and $\ddot{a}$ is undefined. It this case, the degenerated singularity of type III is called sewn (non-isolated) singularity.
We, therefore, obtain \cite{Borowiec:2015qrp}
\begin{equation}
t-t_\text{s}\simeq \pm \frac{1}{2} \frac{d^2 t}{da^2}|_{a=a_{\text{sing}}} (a-a_\text{sing})^2.
\end{equation}
Above formula combine two types of behavior near the freeze singularities in the future
\begin{equation}
a-a_\text{sing}\propto -(t_{\text{sing}}-t)^{1/2} \text{ for } t\rightarrow t_{\text{sing}^{-}}
\end{equation}
and in the past
\begin{equation}
a-a_\text{sing}\propto +(t-t_{\text{sing}})^{1/2} \text{ for } t\rightarrow t_{\text{sing}^{+}}.
\end{equation}
Figure \ref{fig:1} illustrates the behavior of the scale factor in cosmological time in neighborhood of a pole of the potential function. Diagram of $a(t)$ is constructed from the dynamics in two disjoint region $\{a \colon a<a_\text{s}\}$ and $\{a \colon a>a_\text{s}\}$. Figure \ref{fig:6} presents the behavior of the scale factor in the
 cosmological time in a neighborhood of the sudden singularity.

In the model under consideration another type of sewn singularity also appears. It is a composite singularity with two sudden singularities glued at the bounce when $a=a_{\text{min}}$. In this singularity the potential itself is a continuous function while its first derivative has a discontinuity. Therefore, the corresponding dynamical system represents a piece-wise smooth dynamical system.

The problem of $C^0$ metric extension beyond the future Cauchy horizon, when the second derivative of the metric is inextendible, was discussed in work of Jan Sbierski \cite{Sbierski:2015nta}. In the context of FLRW cosmological models, Sbierski's methodology was considered in \cite{Galloway:2016bej}.

\section{Singularities in the Starobinsky model in the Palatini formalism}

In our model, one finds two types of singularities, which are a consequence of the Palatini formalism: the freeze and sudden singularity. The freeze singularity appears when the multiplicative expression $\frac{b}{b+d/2}$, in the Friedmann equation (\ref{friedmann2}), is equal the infinity. So we get a condition for the freeze singularity: $2b+d=0$ which produces a pole in the potential function. It appears that the sudden singularity appears in our model when the multiplicative expression $\frac{b}{b+d/2}$ vanishes. This condition is equivalent to the case $b=0$.

The freeze singularity in our model is a solution of the algebraic equation
 \begin{equation}
 	2b+d=0 \Longrightarrow f(K,\Omega_{\Lambda,0},\Omega_{\gamma})=0
 \end{equation}
or
 \begin{equation}
-3K-\frac{K}{3\Omega_{\gamma}(\Omega_\text{m}+\Omega_{\Lambda,0})\Omega_{\Lambda,0}}+1=0,\label{k}
 \end{equation}
where $K\in [0,\text{ }3)$.

\begin{figure}
	\centering
	\includegraphics[width=0.7\linewidth]{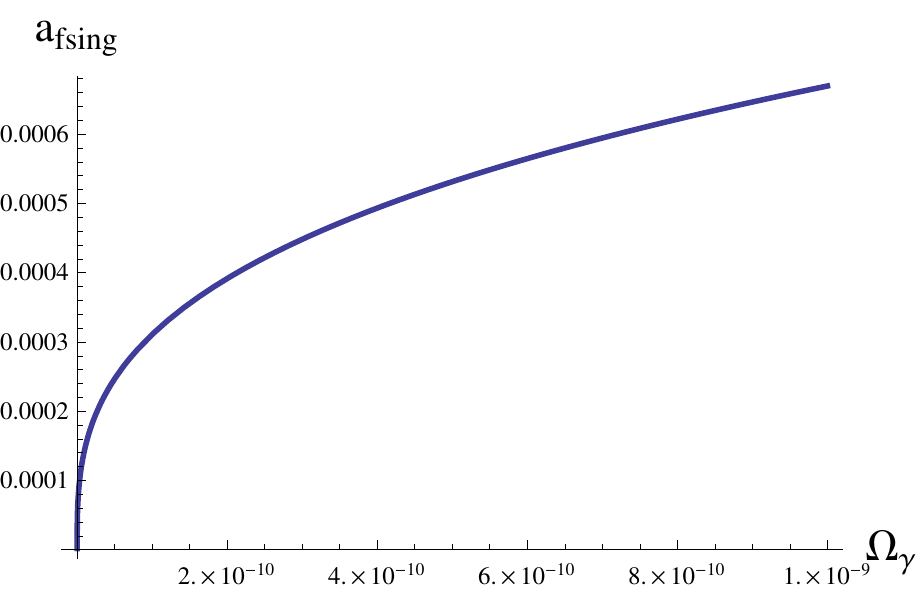}
	\caption{Diagram of the relation between $a_{\text{sing}}$ and positive $\Omega_\gamma$. Note that in the limit $\Omega_\gamma\mapsto 0$ the singularity overlaps with a big-bang singularity.}
	\label{fig:7}
\end{figure}

The solution of the above equation is
 \begin{equation}
 	 K_{\text{freeze}}=\frac{1}{3+\frac{1}{3\Omega_{\gamma}(\Omega_\text{m}+\Omega_{\Lambda,0})\Omega_{\Lambda,0}}}.\label{k2}
 \end{equation}
From equation (\ref{k2}), we can find an expression for a value of the scale factor for the freeze singularity
 \begin{equation}
 	 a_{\text{freeze}}=\left(\frac{1-\Omega_{\Lambda,0}}{8\Omega_{\Lambda,0}+\frac{1}{\Omega_{\gamma}(\Omega_\text{m}+\Omega_{\Lambda,0})}}\right)^\frac{1}{{3}}.
 \end{equation}
The relation between $a_{\text{freeze}}$ and positive $\Omega_\gamma$ is presented in Figure \ref{fig:7}.

\begin{figure}
	\centering
	\includegraphics[width=0.7\linewidth]{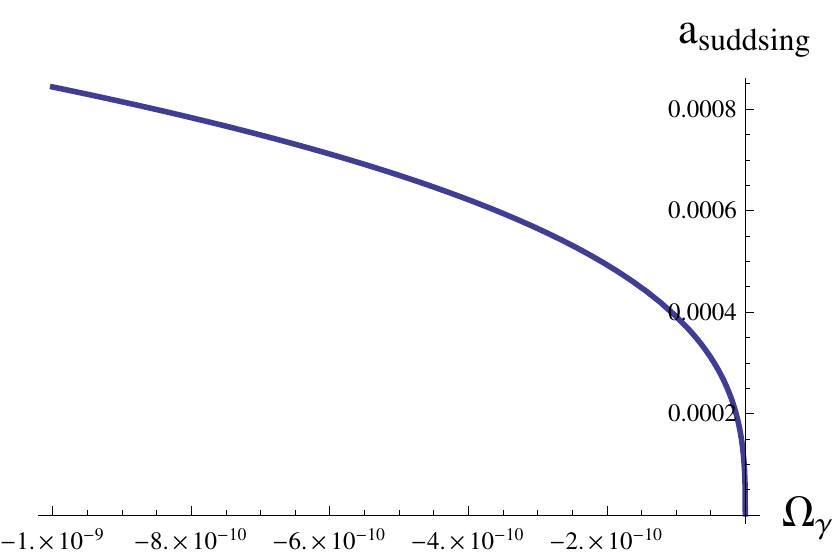}
	\caption{Diagram of the relation between $a_{\text{sing}}$ and negative $\Omega_\gamma$. Note that in the limit $\Omega_\gamma\mapsto 0$ the singularity overlaps with a big-bang singularity.}
	\label{fig:8}
\end{figure}

The sudden singularity appears when $b=0$. This provides the following algebraic equation
\begin{equation}
1+2\Omega_\gamma(\Omega_\text{m,0}a^{-3}+\Omega_{\Lambda,0})(K+1)=0.
\end{equation}
The above equation can be rewritten as
\begin{equation}
1+2\Omega_\gamma(\Omega_{\text{m},0}a^{-3}+4\Omega_{\Lambda,0})=0.\label{sud}
\end{equation}
From the equation (\ref{sud}), we have the formula for the scale factor for sudden singularity
\begin{equation}
a_{\text{sudden}}=\left(-\frac{2\Omega_\text{m,0}}{\frac{1}{\Omega_\gamma}+8\Omega_{\Lambda,0}}\right)^{1/3}.
\end{equation}
which, in fact, becomes a (degenerate) critical point and a bounce at the same time.
The relation between $a_{\text{sing}}$ and negative $\Omega_\gamma$ is presented in Figure \ref{fig:8}.

Let $V=-\frac{ a^2}{2}\left(\Omega_{\gamma}\Omega_{\text{ch}}^2\frac{(K-3)(K+1)}{2b}+\Omega_{\text{ch}}+\Omega_k\right)$. We can rewrite dynamical system (\ref{dota})-(\ref{dotx}) as
\begin{align}
	a' &= x, \label{eq:ds1}\\
	x' &= -\frac{\partial V(a)}{\partial a}, \label{eq:ds2}
\end{align}
where $'\equiv\frac{d}{d\sigma}=\frac{b+\frac{d}{2}}{b}\frac{d}{d\tau}$ is a new parametrization of time.

We can treated the dynamical system (\ref{eq:ds1})-(\ref{eq:ds2}) as a sewn dynamical system \cite{Hrycyna:2008gk,Ellis:2015bag}. In this case, we divide the phase portrait into two parts: the first part is for $a<a_{\text{sing}}$ and the second part is for $a>a_{\text{sing}}$. Both parts are glued along the singularity.

For $a<a_{\text{sing}}$, dynamical system (\ref{eq:ds1})-(\ref{eq:ds2}) can be rewritten to the corresponding form
\begin{align}
a' &= x, \label{eq:ds3}\\
x' &= -\frac{\partial V_1(a)}{\partial a}, \label{eq:ds4}
\end{align}
where $V_1=V(-\eta (a-a_s)+1)$ and $\eta(a)$ notes the Heaviside function.

\begin{figure}
	\centering
	\includegraphics[width=0.7\linewidth]{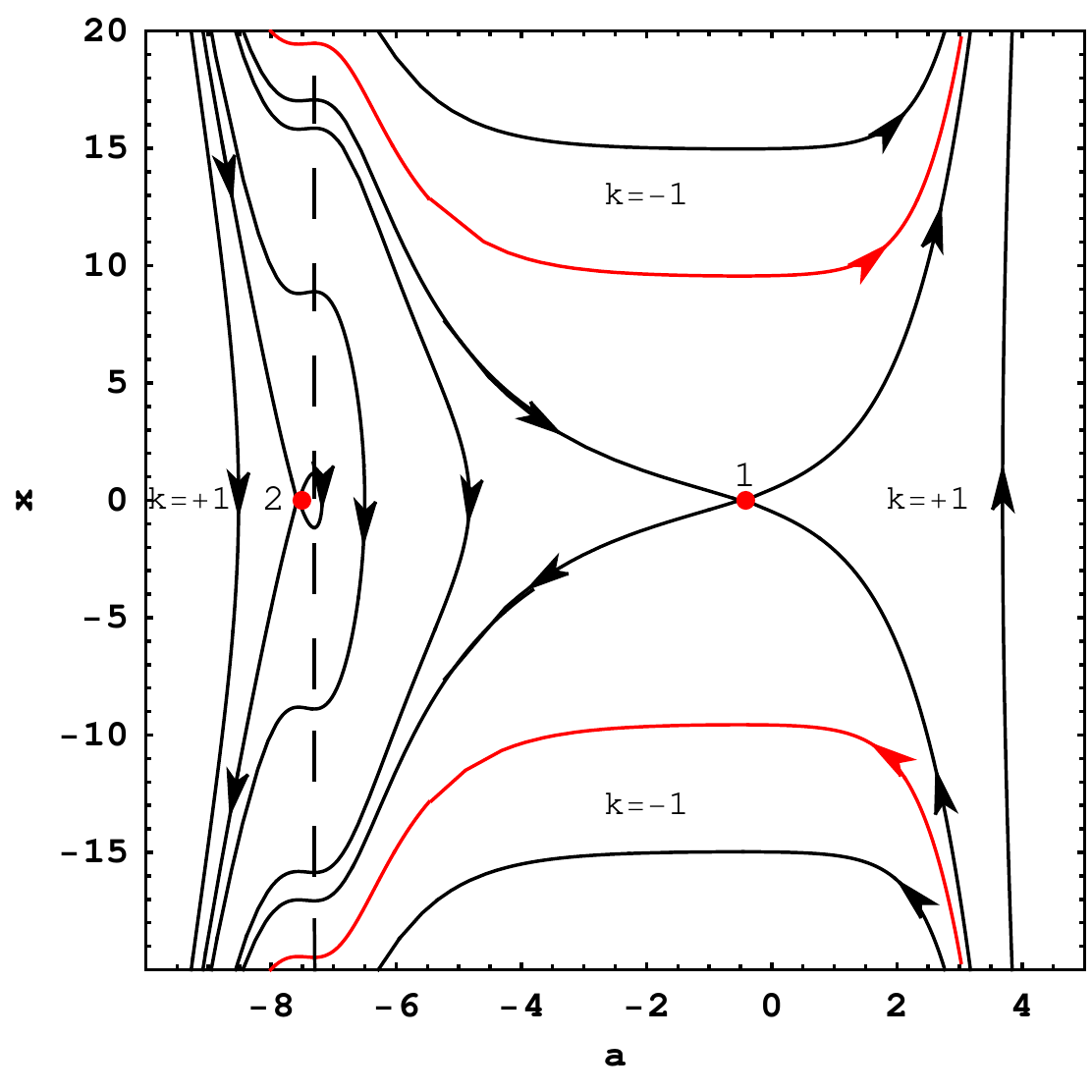}
	\caption{The Figure represents the phase portrait of the system (\ref{eq:ds1}-\ref{eq:ds2}) for positive $\Omega_{\gamma}$. The scale factor $a$ is in the logarithmic scale. The red trajectories represent the spatially flat universe. Trajectories under the top red trajectory and below the bottom red trajectory represent
		models with the negative spatial curvature. Trajectories between the top and bottom red trajectory represent models with the positive spatial curvature. The dashed line $2b+d=0$ corresponds to the freeze singularity. The critical points (1) and (2) present two static Einstein universes. The phase portrait belongs to the class of the sewn dynamical systems \cite{Bautin:1976mt}.}
	\label{fig:2}
\end{figure}

\begin{figure}
	\centering
	\includegraphics[width=0.7\linewidth]{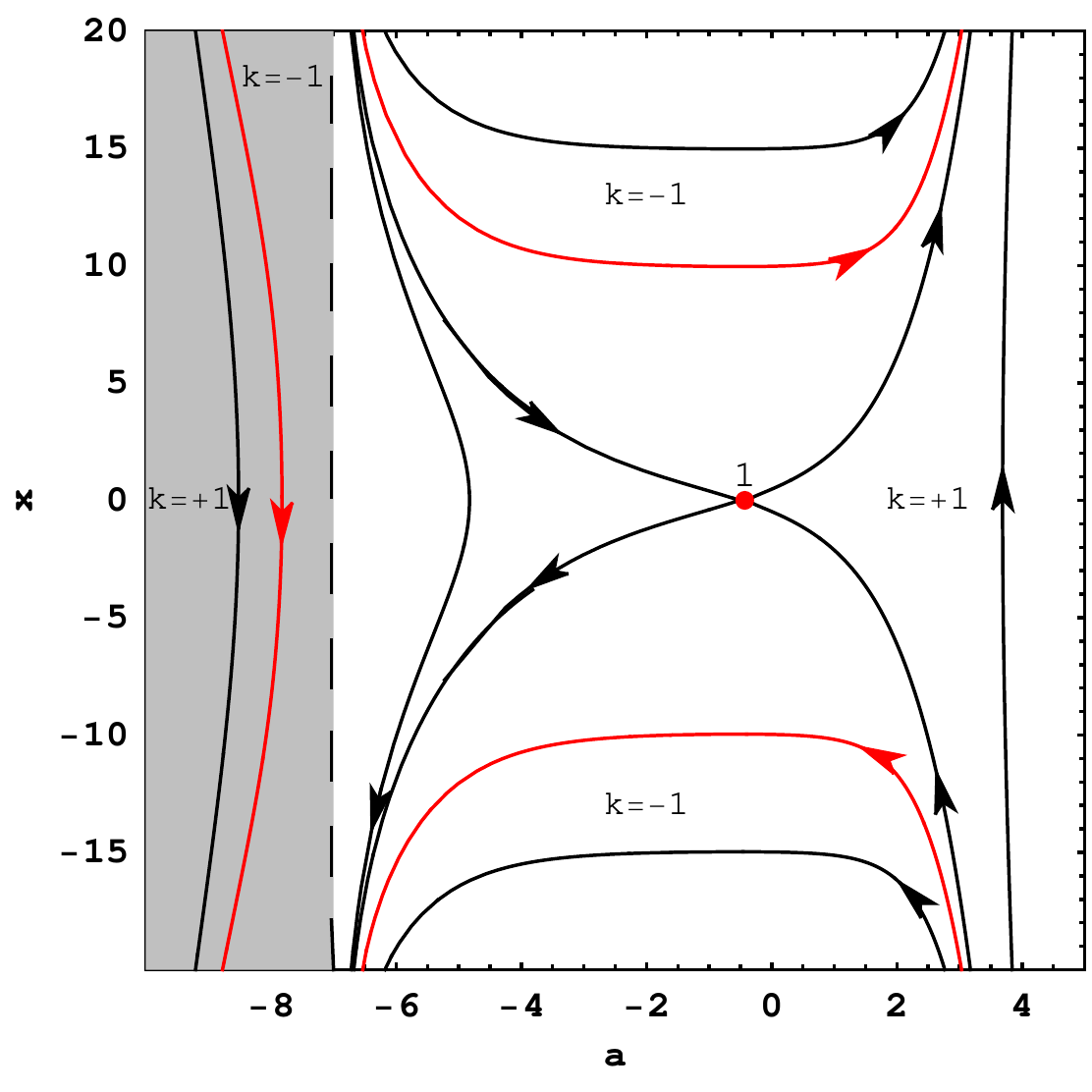}
	\caption{The phase portrait of the system (\ref{eq:ds1}-\ref{eq:ds2}) for negative $\Omega_{\gamma}$. The scale factor $a$ is in logarithmic scale. The red trajectories represent a spatially flat universe. Trajectories under the top red trajectory and below the bottom red trajectory represent models with the negative spatial curvature. Trajectories between the top and bottom red trajectory represent models with the positive spatial curvature. The dashed line $b=0$ corresponds to the sudden singularity. The shaded region represents trajectories with $b<0$. If we assume that $f'(R)>0$ then this region can be removed. The phase portrait possesses the symmetry $\dot{a}\rightarrow-\dot{a}$ and in consequence this singularity presents a bounce. This symmetry can be used to identify the corresponding points on the $b$-line. The critical point (1) presents the static Einstein universe. The phase portrait belongs to the class of the sewn dynamical systems \cite{Bautin:1976mt}.}
	\label{fig:3}
\end{figure}

For $a>a_{\text{sing}}$, in the analogous way, we get the following equations
\begin{align}
a' &= x, \label{eq:ds5}\\
x' &= -\frac{\partial V_2(a)}{\partial a}, \label{eq:ds6}
\end{align}
where $V_2=V \eta (a-a_s)$. The phase portraits, for dynamical system (\ref{eq:ds1})-(\ref{eq:ds2}), are presented in Figure \ref{fig:2} and \ref{fig:3}. Figure \ref{fig:2} provides the phase portrait for positive $\Omega_{\gamma}$ while Figure \ref{fig:3} provides the phase portrait for negative $\Omega_{\gamma}$.

\begin{figure}
	\centering
	\includegraphics[width=0.7\linewidth]{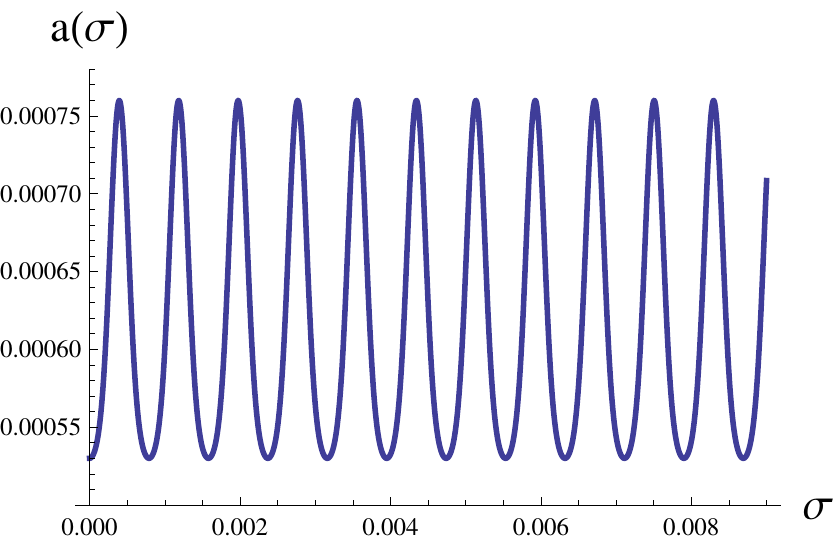}
	\caption{Illustration of the evolution of $a(\sigma)$ for closed orbit which is contained by the homoclinic orbit, where $\sigma=\frac{b}{b+\frac{d}{2}} t$ is a reparametrization of time. We choose s$\times$Mpc/(100$\times$km) as a unit of $\sigma$.}
	\label{fig:10}
\end{figure}

\begin{figure}
	\centering
	\includegraphics[width=0.7\linewidth]{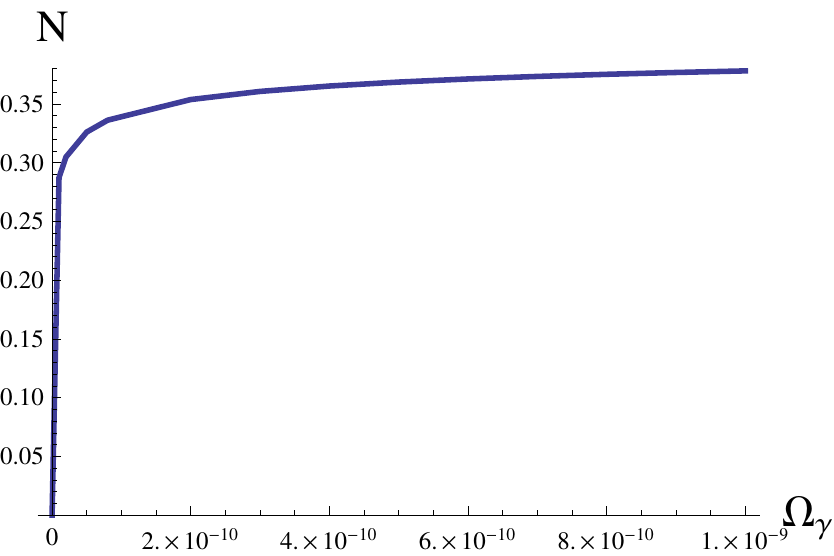}
	\caption{Diagram of the relation between positive $\Omega_{\gamma}$ and the approximate number of e-foldings $N=H_{\text{init}} (t_{\text{fin}} -t_{\text{init}})$ from $t_\text{init}$ to $t_\text{fin}$.}
	\label{fig:9}
\end{figure}

In Figure \ref{fig:2} there are two critical points labelled `1' and `2' at the finite domain. They are both the saddles. These critical points correspond to a maximum of the potential function. The saddle `2' possesses the homoclinic closed orbit starting from it and returning to it. This orbit represents an emerging universe from the static Einstein universe and approaching to it again. During the evolution this universe (orbit) goes two times through the freeze singularity. The region bounded by the homoclinic orbit contains closed orbits representing the oscillating universes. Diagram of the evolution of scale factor for closed orbit is presented by Figure \ref{fig:10}. It is also interesting that trajectories in neighborhood of straight vertical line of freeze singularities undergo short time inflation $x = \text{const}$. The characteristic number of e-foldings from $t_\text{init}$ to $t_\text{fin}$ of this inflation period $N=H_{\text{init}} (t_{\text{fin}} -t_{\text{init}})$ (see formula (3.13) in \cite{DeFelice:2010aj}) with respect to $\Omega_{\gamma}$ is shown in Figure \ref{fig:9}. This figure illustrates the number of e-foldings is too small for to obtain the inflation effect.


\section{Observations}

In this paper we perform statistical analysis using the following astronomical observations: observations of 580 supernovae of type Ia, BAO, measurements of $H(z)$ for galaxies, Alcock-Paczy{\'n}ski test, measurements of CMB and lensing by Planck and low $\ell$ by WMAP.

The likelihood function for observations of supernovae of type Ia \cite{Suzuki:2011hu} is given by the following expression
\begin{equation}
	\ln L_{\text{SNIa}} = -\frac{1}{2} [A - B^2/C + \ln(C/(2 \pi))],
\end{equation}
where $A=
(\mathbf{\mu}^{\text{obs}}-\mathbf{\mu}^{\text{th}})\mathbb{C}^{-1}(\mathbf{\mu}^{\text{obs}}-\mathbf{\mu}^{\text{th}})$,
$B=
\mathbb{C}^{-1}(\mathbf{\mu}^{\text{obs}}-\mathbf{\mu}^{\text{th}})$, $C=\text{Tr} \mathbb{C}^{-1}$ and $\mathbb{C}$ is a covariance matrix for observations of supernovae of type Ia. The distance modulus is defined by the formula $\mu^{\text{obs}}=m-M$ (where $m$ is the apparent magnitude and $M$ is the absolute magnitude of observations of supernovae of type Ia) and $\mu^{\text{th}} = 5
\log_{10} D_L +25$ (where the luminosity distance is $D_L= c(1+z)
\int_{0}^{z} \frac{d z'}{H(z)}$).

BAO observations such as Sloan Digital Sky Survey Release 7 (SDSS DR7) dataset at $z = 0.275$ \cite{Percival:2009xn}, 6dF Galaxy Redshift Survey measurements at redshift $z = 0.1$ \cite{Beutler:2011hx}, and WiggleZ measurements at redshift $z = 0.44, 0.60, 0.73$
\cite{Blake:2012pj} have the following likelihood function
\begin{equation}
	\ln L_{\text{BAO}} = -
	 \frac{1}{2}\left(\mathbf{d}^{\text{obs}}-\frac{r_s(z_d)}{D_V(\mathbf{z})}\right)\mathbb{C}^{-1}\left(\mathbf{d}^{\text{obs}}-\frac{r_s(z_d)}{D_V(\mathbf{z})}\right),
\end{equation}
where $r_s(z_d)$ is the sound horizon at the drag epoch
\cite{Hu:1995en,Eisenstein:1997ik}.

For the Alcock-Paczynski test \cite{Alcock:1979mp,Lopez-Corredoira:2013lca} we used the following expression for the likelihood function
\begin{equation}
	\ln L_{AP} = - \frac{1}{2} \sum_i \frac{\left(
		AP^{th}(z_i)-AP^{obs}(z_i) \right)^2}{\sigma^2}.
\end{equation}
where $AP(z)^{\text{th}} \equiv \frac{H(z)}{z} \int_{0}^{z}
\frac{dz'}{H(z')}$ and $AP(z_i)^{\text{obs}}$ are observational
data
\cite{Sutter:2012tf,Blake:2011ep,Ross:2006me,Marinoni:2010yoa,daAngela:2005gk,Outram:2003ew,Anderson:2012sa,Paris:2012iw,Schneider:2010hm}.

The likelihood function for measurements of the Hubble parameter $H(z)$ of galaxies from
\cite{Simon:2004tf,Stern:2009ep,Moresco:2012jh} is given by the expression
\begin{equation}\label{hz}
	\ln L_{H(z)} = -\frac{1}{2} \sum_{i=1}^{N} \left
	(\frac{H(z_i)^{\text{obs}}-H(z_i)^{\text{th}}}{\sigma_i
	}\right)^2.
\end{equation}

In this paper, we use the likelihood function for observations of CMB \cite{Ade:2015rim} and lensing by Planck, and low-$\ell$ polarization from the WMAP (WP) in the following form
\begin{equation}
\ln L_{\text{CMB}+\text{lensing}} = - \frac{1}{2} (\mathbf{x}^{\text{th}}-\mathbf{x}^{\text{obs}})
\mathbb{C}^{-1} (\mathbf{x}^{\text{th}}-\mathbf{x}^{\text{obs}}),
\end{equation}
where $\mathbb{C}$ is the covariance matrix with the errors, $\mathbf{x}$ is a vector of the acoustic scale $l_{A}$, the shift parameter $R$ and $\Omega_{b}h^2$ where
\begin{align}
l_A &= \frac{\pi}{r_s(z^{*})} c \int_{0}^{z^{*}} \frac{dz'}{H(z')} \\
R &= \sqrt{\Omega_{\text{m},0} H_0^2} \int_{0}^{z^{*}} \frac{dz'}{H(z')},
\end{align}
where $z^{*}$ is the redshift of the epoch of the recombination \cite{Hu:1995en}.

The total likelihood function is expressed in the following form
\begin{equation}
	L_{\text{tot}} = L_{\text{SNIa}} L_{\text{BAO}} L_{\text{AP}}
	L_{H(z)} L_{\text{CMB+lensing}}.
\end{equation}

To estimate model parameters, we use our own code CosmoDarkBox. The Metropolis-Hastings algorithm \cite{Metropolis:1953am, Hastings:1970aa} is used in this code.

\begin{figure}[ht]
	\centering
	\includegraphics[scale=1]{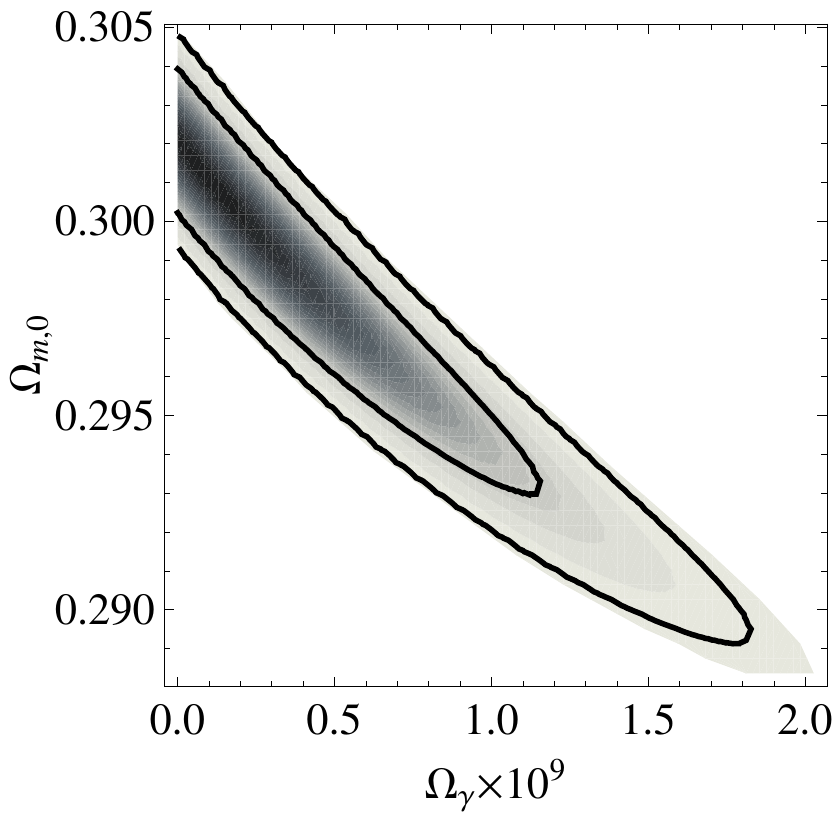}
	\caption{The intersection of the likelihood functions of two model parameters ($\Omega_{\gamma}$, $\Omega_{m,0}$) with the marked $68\%$ and $95\%$ confidence levels.}
	\label{fig:4}
\end{figure}

\begin{figure}[ht]
	\centering
	\includegraphics[scale=1]{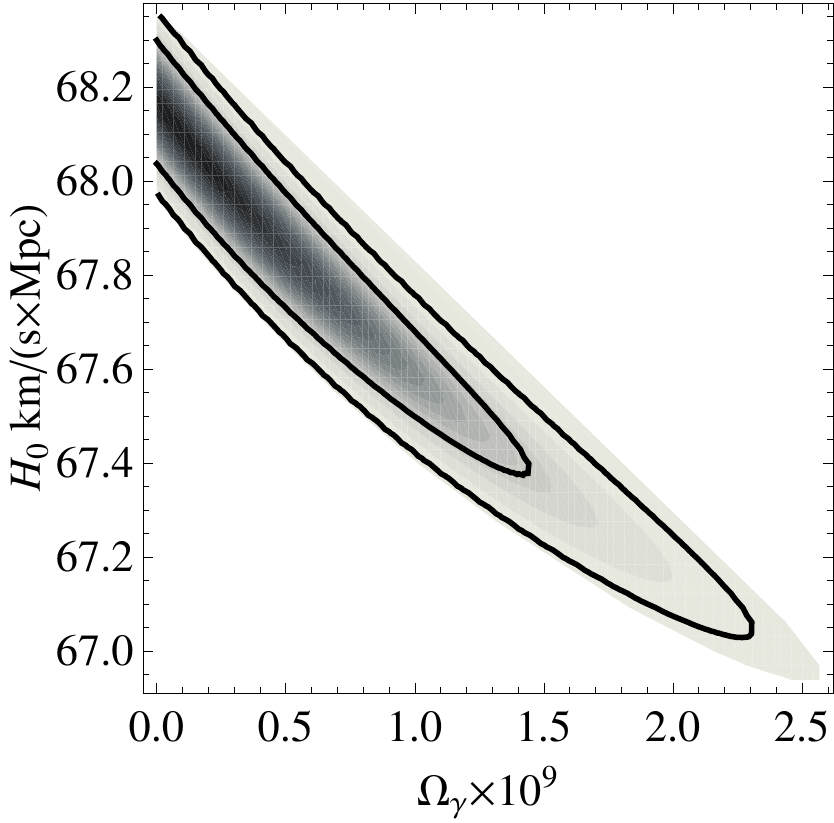}
	\caption{The intersection of the likelihood functions of two model parameters ($\Omega_{\gamma}$, $H_0$) with the marked $68\%$ and $95\%$ confidence levels.}
	\label{fig:5}
\end{figure}

Table \ref{table:1} completes the values of parameters for the best fit with errors. Figure~\ref{fig:4} and Figure~\ref{fig:5} show the intersection of a likelihood function with the $68\%$ and $95\%$ confidence level projections on the planes ($\Omega_{\gamma}$, $\Omega_{\text{m},0}$) and ($\Omega_{\gamma}$, $H_0$).

In this paper, we use Bayesian information criterion (BIC) \cite{Schwarz:1978ed, Kass:1995bf}, for comparison our model with the $\Lambda$CDM model. The expression for BIC is defined as
\begin{equation}
	\text{BIC}=\chi^2+j \ln n,
\end{equation}
where $\chi^2$ is the value of $\chi^2$ in the best fit, $j$ is the number of model parameters (our model has three parameters, $\Lambda$CDM model has two parameters) and $n$ is number of data points ($n=625$), which are used in the estimation.

For our model, the value of BIC is equal 135.668 and for the $\Lambda$CDM model BIC$_{\Lambda\text{CDM}}=129.261$. So $\Delta$BIC=BIC-BIC$_{\Lambda\text{CDM}}$ is equal 6.407. The evidence for the model is strong \cite{Kass:1995bf} if $\Delta$BIC is more than 6. So, in comparison to our model, the evidence in favor of the $\Lambda$CDM model is strong, but we cannot absolutely reject our model.

\begin{table}
	\caption{The best fit and errors for the estimated model for the positive $\Omega_\gamma$ with $\Omega_{\text{m},0}$ from the interval $(0.27, 0.33)$, $\Omega_{\gamma}$ from the interval $(0.0, 2.6\times 10^{-9})$ and $H_0$ from the interval (66.0 (km/(s Mpc)), 70.0 (km/(s Mpc))). $\Omega_{\text{b},0}$ is assumed as 0.048468. The redshift of matter-radiation equality is assumed as 3395. $H_0$, in the table, is expressed in km/(s Mpc). The value of reduced $\chi^2$ of the best fit of our model is equal 0.187066 (for the $\Lambda$CDM model 0.186814).}
	\label{table:1}
\begin{center}
		\begin{tabular}{llll} \hline
			parameter & best fit & $ 68\% $ CL & $ 95\% $ CL  \\ \hline \hline
			$H_0$ & 68.10 & $\begin{array}{c}
			+1.07 \\ -1.24
			\end{array}$ & $\begin{array}{c}
			+1.55 \\ -1.82
			\end{array}$ \\ \hline
			$\Omega_{\text{m},0}$ & 0.3011 & $\begin{array}{c}
			+0.0145 \\ -0.0138
			\end{array}$ & $\begin{array}{c}
			+0.0217 \\ -0.0201
			\end{array}$ \\ \hline
			$\Omega_{\gamma}$ & $9.70\times 10^{-11}$ & $\begin{array}{c}
			+1.3480\times 10^{-9} \\ -9.70\times 10^{-11}
			\end{array}$ & $\begin{array}{c}
			+2.2143\times 10^{-9} \\ -9.70\times 10^{-11}
			\end{array}$ \\ \hline
		\end{tabular}
	\end{center}
\end{table}

\section{Conclusions}

In this paper, we demonstrated that evolution of the Starobinsky model with a quadratic term $R^2$ gives rise to the description of dynamics in terms of piecewise-smooth dynamical systems (PWS), i.e., systems whose the phase space is partitioned into different regions, each of them associated to a different smooth functional form of the system of a Newtonian type. Different regions of the phase space correspond to different forms of the potential separated by singularities of the type of poles.

Our idea was to obtain the inflation as an endogenous effect of the dynamics in the Palatini formalism. While the effect of inflation appears in the model under consideration the number of sufficient e-folds is not achieved and the additional effect of amplification is required. Note that this type of inflation is a realisation of the idea of singular inflation \cite{Barrow:2015ora,Bamba:2016ngh,Nojiri:2015wsa,Odintsov:2015gba}. In our model inflation is driven by the freeze degenerated singularity (the extension of type III isolated singularity).

We show that dynamics of the model can be analyzed in terms of two-dimensional dynamical systems of the Newtonian type. In this approach, in the diagram of the potential of a fictitious particle mimicking, the evolution of the universe contain all information which are needed for investigation of singularities in the model. Note that they are not isolated singularities which were classified into five types but rather double singularities glued in one point of the evolution at $a=a_\text{sing}$. Appearance of such types of singularities is typical for piece-wise smooth dynamics describing the model evolution. We called this type of sewn singularities in analogy to sewn dynamical systems \cite{Zhusubaliyev:2003bc, Leine:2004db}.

We investigated the model with $f(\hat{R})= \hat{R} +\gamma \hat{R}^2$, where $\gamma$ assumes the positive or negative values. While the dynamics of this class of models depend crucially on the sign of the parameter $\gamma$ in the early universe for the late time we obtain the behavior consistent with the $\Lambda$CDM model.

Note that in the model with positive $\gamma$, the phase space is a sum of two disjoint domains which boundary represents the double freeze singularity (cf. Figure \ref{fig:2}). In the first domain the evolution starts from big-bang followed by the deceleration phase then changes to acceleration (early acceleration $\equiv$ inflation) after reaching a maximum of the potential function. In the second domain, on the right from the vertical line of the freeze singularity, the universe decelerates and after reaching another maximum starts to accelerate again. This last eternal acceleration correspond to the present day epoch called dark energy domination epoch. Two phases of deceleration and two phases of acceleration are key ingredients of our model.
While the first phase models a transition from the matter domination epoch to the inflation the second phase models a transition from the second matter dominated epoch toward the present day acceleration.

As De Felice and Tsujikawa have noted \cite[p.~24]{DeFelice:2010aj} the applications of  $f(R)$ theories
should be focused on construct of viable cosmological models, for which a sequence of radiation, matter and accelerating epochs is realized. All these epochs are also presented in the model under consideration but, for negative $\gamma$ (negative squared $M^2$ for the scalar field), some difficulties appear in the interpretation of the phase space domain $\{a\colon a<a_\text{sing}\}$. The size of this domain will depend on the value of the parameter $\Omega_\gamma$ and this domain vanishes as we are going toward $\Omega_\gamma$ equal zero.

On the other hand it is well know that violation of condition $f''_{\hat{R}\hat{R}}>0$ gives rise to the negative values of  $M^2$. We do not assume this condition but we require that $f'_{\hat{R}}>0$ to avoid the appearance of ghosts (see section 7.4 in \cite{DeFelice:2010aj}). In our case, statistical analysis more favors model with $f'_{\hat{R}}>0$ ($\Omega_\gamma>0$) than with $f'_{\hat{R}}<0$ ($\Omega_\gamma<0$). In other words, statistical analysis more favors the case without ghosts.

In order to obtain deeper insight into the model we have also performed  complementary investigations  in the Einstein frame. In this case we obtain that the model is reduced to the FRW cosmological model with the selfinteracting scalar field  and the vanishing part of the kinetic energy. Therefore from the Palatini formulation we obtain directly the form of the potential and the (implicit) functional dependence between the scalar field  and  the scale factor. Moreover, we obtain the parametrization of the decaying cosmological constant.

Due to time dependent cosmological constant the model evolution can be described in terms of an interaction between the matter and the decaying lambda terms. We study how the energy is transferred between sectors and how the standard scaling relation for matter is modified.

We pointed out that the consideration of the Starobinsky model in the Einstein frame gives rise to  new interesting properties from the cosmological point of view similarly as the original (metric) Starobinsky model is very important  for the  explanation of the inflation. The model under the consideration gives rise analogously  to the running cosmological term. This fact seems to be interesting in the context of an explanation of the cosmological constant problem.

Detailed conclusions coming from our analysis are the following:
\begin{itemize}

\item We show that the interaction between two sectors: the matter and the decaying vacuum, appears naturally   in the Einstein frame. For the model formulated in the Jordan frame this interaction is absent.

\item The inflation appears in our model formulated in the Einstein frame, when the parameter $\gamma$ is close to zero and the density of matter is negligible in comparison to $\bar\rho_\Phi$.

\item In our model in the Einstein frame, the potential $\bar U(\Phi)$ has the same shape like the Starobinsky potential and has the minimum for $\Phi=1+8\gamma\lambda$.

\item While the freeze double singularities appear in our model in the Jordan frame there is no such singularities in the dynamics of the model in the Einstein frame.
	
\item If $\Omega_\gamma$ is small, then $a_{\text{sing}}=\left(-\frac{2\Omega_\text{m,0}}{\frac{1}{\Omega_\gamma}+8\Omega_{\Lambda,0}}\right)^{1/3}$ for negative $\Omega_\gamma$ and $a_{\text{sing}}=\left(\frac{1-\Omega_{\Lambda,0}}{8\Omega_{\Lambda,0}+\frac{1}{\Omega_{\gamma}(\Omega_\text{m}+\Omega_{\Lambda,0})}}\right)^\frac{1}{{3}}$ for positive $\Omega_\gamma$. These values defines the natural scale at which singularities appear in the model under consideration with the negative or positive value of $\gamma$ parameter. It seems to be natural to identify this scale with a cut off at which the model can be treated as some kind of effective theory.

\item In both cases of negative and positive $\gamma $ one deals with a finite scale factor singularity. For negative $\gamma $ it is a double sudden singularity which meets the future singularity of contracting model before the bounce with the initial singularity in the expanding model. The sewn evolutionary scenarios reveal the presence of bounce during the cosmic evolution.

\item In the context of the Starobinsky model in the Palatini formalism we found a new type of double singularity beyond the well-known classification of isolated singularities.

\item The phase portrait for the model with a positive value of $\gamma$ is equivalent to the phase portrait of the $\Lambda$CDM model (following the dynamical system theory \cite{Perko:2001de} equivalence assumes the form of topological equivalence establish by homeomorphism). There is only a quantitative difference related with
the presence of the non-isolated freeze singularity.
The scale of appearance of this type singularity can be also estimated and in terms of redshift: $z_\text{freeze}=\Omega_\gamma^{-1/3}$.

\item We estimated the model parameters using astronomical data and conclude that positive $\Omega_\gamma$ is favored by the best fit value; still the model without $\hat{R}^2$ term is statistically admitted.
\end{itemize}

In our model, the best fit value of $\Omega_\gamma$ is equal $9.70\times 10^{-11}$ and positive $\Omega_\gamma$ parameter belongs to the interval $(0,\text{ }2.2143\times 10^{-9})$ at 2-$\sigma$ level. This mean that the positive value of $\Omega_\gamma$ is more favored by astronomical data than the negative value of $\Omega_\gamma$. The difference between values of BIC for our model and the $\Lambda$CDM model is equal 6.407. So, in comparison to our model, the evidence in favor of the $\Lambda$CDM model is strong. But one cannot absolutely reject the model.

{\it Note added in proof.} After completing the paper we found a paper by Faraoni and Cardini where freeze singularities have been analyzed in a different context, both from point particle and cosmological perspectives \cite{Faraoni:2016lhd}.

\begin{acknowledgements}
AB and MS would like to thank Salvatore Capozziello for encouraging us to consider this model in different frames.
The work has been supported by the Polish National Science Centre (NCN), project DEC-2013/09/B/ ST2/03455.
\end{acknowledgements}

\providecommand{\href}[2]{#2}\begingroup\raggedright\endgroup
\end{document}